%% file: nodeless.tex
\documentclass[aps,reprint,superscriptaddress]{revtex4-1}

\usepackage{hyperref}
\hypersetup{
  colorlinks   = true, 
  urlcolor     = blue, 
  linkcolor    = blue, 
  citecolor   = blue
}
\usepackage{amsmath}
\usepackage{amssymb}

\usepackage[caption=false]{subfig}
\usepackage{tikz}
\definecolor{mtwo}{RGB}{19, 128, 213}
\definecolor{mone}{RGB}{5, 163, 201}
\definecolor{zero}{RGB}{45, 183, 163}
\definecolor{one}{RGB}{100, 187, 130}
\definecolor{two}{RGB}{150, 190, 110}
\definecolor{three}{RGB}{210, 186, 88}
\definecolor{four}{RGB}{251, 187, 64}
\definecolor{five}{RGB}{248, 215, 39}
\definecolor{six}{RGB}{248, 250, 13}

\usepackage{soul} 
\begin{document}

\title{Topological Phases in Nodeless Tetragonal Superconductors}

\author{S. Varona}
\affiliation{Departamento de F\'{\i}sica Te\'orica, Universidad Complutense, 28040 Madrid, Spain}
\author{L. Ortiz}
\affiliation{Departamento de F\'{\i}sica Te\'orica, Universidad Complutense, 28040 Madrid, Spain}
\author{O. Viyuela}
\affiliation{Department of Physics, Harvard University, Cambridge, MA 02318, USA}
\affiliation{Department of Physics, Massachusetts Institute of Technology, Cambridge, MA 02139, USA}
\author{M.A. Martin-Delgado}
\affiliation{Departamento de F\'{\i}sica Te\'orica, Universidad Complutense, 28040 Madrid, Spain}



\begin{abstract}
We compute the topological phase diagram of 2D tetragonal superconductors for the only possible nodeless pairing channels compatible with that crystal symmetry. Subject to a Zeeman field and spin-orbit coupling, we demonstrate that these superconductors show surprising topological features: non-trivial high Chern numbers, massive edge states, and zero-energy modes out of high symmetry points, even though the edge states remain topologically protected. Interestingly, one of these pairing symmetries, $d+id$, has been proposed to describe materials such as water-intercalated sodium cobaltates, bilayer silicene or highly doped monolayer graphene, which opens the way for further applications of our results. 
\end{abstract}


\maketitle

\section{\label{intro}Introduction}

Topological phases of matter have acquired a prominent role in condensed matter physics and quantum computation since the most outstanding examples were proposed \cite{PhysRevLett.49.405,HALDANE1983464,doi:10.1142/S0217979292000840}.  Developments on 2D superconductors have led to great interest in investigating novel quantum phenomena. There is an increasing body of experimental evidence that superconductivity can be remarkably robust in the two-dimensional (2D) limit, both in conventional and in high-$T_c$ superconductors \cite{book_martin-delgado}.

Particularly, topological superconductors have gained additional interest since they host Majorana fermions, that could be used as building blocks of future topological quantum computers \cite{RevModPhys.80.1083,RepAlicea,PhysRevX.7.031048}. 
The odd superconducting pairing mechanism required to display topological features has not being found in Nature, except for the B phase of superfluid $^3$He  \cite{Volovik1988} and probably ruthenates \cite{nature1994}. However, it can be synthesized in different experimental platforms: topological insulators \cite{PhysRevLett.100.096407,PhysRevLett.114.017001} and semiconductors \cite{Mourik1003,doi:10.1021/nl303758w,nat_charlie_markus}  proximity coupled to superconductors, diluted magnetic impurities in superconducting lead \cite{Yazdani}, iron-based superconductors \cite{Wang2017}, quantum anomalous Hall insulator \cite{He2017}, etc. These experiments make use of underlying superconductors to eventually observe Majorana fermions. 
Most of these proposals  use conventional superconductors \cite{PhysRevLett.100.096407, PhysRevB.81.125318, PhysRevLett.105.077001}, although there are some for high-$T_c$ superconductors \cite{Zareapour2012,PhysRevB.94.134518,PhysRevB.91.235143,1803.08545} as well.

Likewise, describing the more exotic high-$T_c$ superconductors has motivated a lot of research  since they were experimentally discovered \cite{Bednorz1986}.  Shortly after, the resonating valence bond theory came up as the first theoretical proposal to describe these materials \cite{ANDERSON1196}. Despite the microscopic origin of high-$T_c$ superconductors is still unknown, these materials were shown to present $d$-wave pairing symmetry, tetragonal crystal symmetry and singlet pairing \cite{RevModPhys.72.969}.  The question of  understanding  what generic properties might be expected of singlet-paired tetragonal superconductors gave rise to a systematic  symmetry classification of all possible  pairing channels \cite {WengerPhysRevB.47.5977}.
 
 Singlet pairing in tetragonal superconductors can take the well-known forms of $s$-wave and $d$-wave states. However, it is also possible to have alternative singlet pairings which are compatible with the symmetry group of tetragonal crystals $D_{4h}$. These pairings might have some interesting implications regarding topological phases and the appearance of Majorana fermions. Of particular interest are the four different mixed nodeless pairings that can be formed by summing 1D irreducible representations of $D_{4h}$. One instance of these nodeless pairings is $d+id$ pairing \cite{PhysRevB.60.4245,PhysRevB.62.99,PhysRevB.61.10267,Sato2010,doi:10.1063/1.4961462,PhysRevB.96.014521}, which has potential applications and has been proposed in materials such as water-intercalated sodium cobaltates, bilayer silicene, epitaxial bilayer films of bismuth and nickel \cite{PhysRevLett.111.097001, PhysRevLett.111.066804,doi:10.1063/1.4961462,PhysRevB.94.064519, PhysRevLett.100.217002,Gong2017}, or highly doped monolayer graphene \cite{natphys8}. Thus, in this paper we wonder about the role of these less-studied nodeless pairing channels in the search for distinct topological phases of matter.  

Superconductors have a particularly rich topological behavior in the presence of spin-orbit coupling (SOC) and a Zeeman field \cite{PhysRevB.96.224512,PhysRevB.94.184517}.  When also considering $d$-wave superconductors, the presence of nodal lines gives rise to bulk states at zero energy and cause two main undesired effects: (i) The Chern number is ill-defined since the gap closes at the nodal points. (ii) Although the parity of the Chern number is a well-defined topological invariant and Majorana states are topologically protected, these may interact with nodal states in disordered systems \cite{ PhysRevLett.105.217001,PhysRevB.97.064501}. 
Remarkably, some of the tetragonal pairings previosly mentioned are nodeless. Thus, there are no nearby zero energy modes that may spoil the topological protection of the edge states. Since the gap does not close, the Chern number is well defined and related to the number of edge states via the bulk-edge correspondence \cite{Fukui2012,RevModPhysRyu,PhysRevB.96.184516}.

In this work we analyse the four possible  nodeless superconducting pairings  (compatible with $D_{4h}$ symmetry) thoroughly, using the Chern number and the bulk-edge correspondence, finding the following results: (i) Despite the Chern number coincides with the number of edge states, the number of zero-energy modes is not necessarily the same as the Chern number.  (ii) Massive edge states can be found for $d+id$ pairing when the upper band Chern number takes non-zero values. (iii)
 $d+id$ also presents zero energy modes out of the time-reversal-invariant momenta of the Brillouin Zone.  This is quite remarkable, since in most cases the zero modes are naturally placed a these highly symmetric points. We explicitly show how the edge states satisfy the bulk-edge correspondence  and test their robustness to weak disorder perturbations. Edge states appearing for $d+id$ pairing have been previously observed \cite{PhysRevB.60.4245,PhysRevB.62.99,PhysRevB.61.10267,Sato2010,doi:10.1063/1.4961462}. However, a detailed construction and explanation of their existence was still missing, to best of our knowledge. A complementary analysis of the $D_{6h}$ group would allow us to extend the results presented here for the case of hexagonal lattices. Even though the results would be qualitatively different, the analytic procedure and numerical methods developed in this work still hold.

The article is organized as follows. In Sec. \ref{sec:formalism} we introduce the four different pairings we want to study with tetragonal symmetry $D_{4h}$. In Sec. \ref{sec:phase_diag}, we study their induced topological phases. In Sec. \ref{sec:edge_states} we compute the topological edge states and zero-energy modes. Sec. \ref{sec:conclusions} is devoted to conclusions.


\section{Formalism} \label{sec:formalism}

In this section, we introduce a Hamiltonian on a square lattice to  study the topological phases arising from 2D singlet superconductors with tetragonal $D_{4h}$ symmetry. The necessary ingredients to have topological behavior are SOC, Zeeman field and superconducting pairing. Concretely, we analyze all possible nodeless pairing compatible with tetragonal symmetry. As it is mentioned in the introduction, nodeless pairings are particularly interesting. The Hamiltonian for these systems reads
\begin{equation}\label{eq:H_sum}
{\cal H}=\frac{1}{2}\sum_{\boldsymbol{k},\sigma,\sigma'}\left(c_{\boldsymbol{k},\sigma}^{\dagger},c_{-\boldsymbol{k},\sigma}\right)H\left(\boldsymbol{k}\right)\left(\begin{array}{c}
c_{\boldsymbol{k},\sigma'}\\
c_{-\boldsymbol{k},\sigma'}^{\dagger}
\end{array}\right), 
\end{equation}
where
\begin{align} \label{eq:H_matrix}
&H\left(\boldsymbol{k}\right)=\nonumber\\
&\left(\begin{array}{cc}
\epsilon\left(\boldsymbol{k}\right)-V\sigma_{z}+\boldsymbol{g}\left(\boldsymbol{k}\right)\cdot\boldsymbol{\sigma} & i\Delta\left(\boldsymbol{k}\right)\sigma_{y}\\
-i\Delta^{*}\left(\boldsymbol{k}\right)\sigma_{y} & -\epsilon\left(\boldsymbol{k}\right)+V\sigma_{z}+\boldsymbol{g}\left(\boldsymbol{k}\right)\cdot\boldsymbol{\sigma}^{*}
\end{array}\right),
\end{align}
with $\epsilon\left(\boldsymbol{k}\right)=-2t\left(\cos k_{x}+\cos k_{y}\right)-\mu$,
$\boldsymbol{g}\left(\boldsymbol{k}\right)=\alpha\left(\sin k_{y},-\sin k_{x},0\right)$ is the SOC, $V$ is the Zeeman field, $\Delta\left(\boldsymbol{k}\right)$ is the superconducting pairing and $\boldsymbol{\sigma}$ are the Pauli matrices. $\Delta\left(\boldsymbol{k}\right)$ is an even function $\Delta\left(-\boldsymbol{k}\right)=\Delta\left(\boldsymbol{k}\right)$ as required by singlet pairing. The Hamiltonian is particle-hole symmetric, i.e., $\Gamma  H\left(\boldsymbol{k}\right) \Gamma^\dagger = - H^*\left(-\boldsymbol{k}\right)$, with $\Gamma=\sigma_x\otimes\mathbb{I}$.
The energy bands for this Hamiltonian take the form
\begin{alignat}{1}
 E_{\pm}^2\left(\boldsymbol{k}\right)&=\epsilon^{2}\left(\boldsymbol{k}\right)+\left|\boldsymbol{g}\left(\boldsymbol{k}\right)\right|^{2}+V^{2}+\left|\Delta\left(\boldsymbol{k}\right)\right|^{2}\nonumber \\
 &\pm2\sqrt{\epsilon^{2}\left(\boldsymbol{k}\right)\left|\boldsymbol{g}\left(\boldsymbol{k}\right)\right|^{2}+\epsilon^{2}\left(\boldsymbol{k}\right)V^{2}+\left|\Delta\left(\boldsymbol{k}\right)\right|^{2}V^{2}},\label{eq:Ebands}
\end{alignat}
where $E_+$ is the upper band and $E_-$ the lower. Due to particle-hole symmetry we also have the hole-like solutions $-E_+$ and $-E_-$.  It is important to obtain  the conditions for which the energy gap closes, since these will signal a topological phase transition. From Eq.\ \eqref{eq:Ebands} one can obtain the conditions that must be satisfied for the lower band gap, $E_-\left(\boldsymbol{k}\right)$, to close
\begin{gather}
\epsilon^{2}\left(\boldsymbol{k}\right)+\left|\Delta\left(\boldsymbol{k}\right)\right|^{2}-V^{2}-\left|\boldsymbol{g}\left(\boldsymbol{k}\right)\right|^{2}=0,\label{eq: gap closing 1}\\ 
\left|\Delta\left(\boldsymbol{k}\right)\right|\left|\boldsymbol{g}\left(\boldsymbol{k}\right)\right|=0.\label{eq: gap closing 2}
\end{gather}
\begin{figure*}
\subfloat[\label{fig:splusis_phase}]{%
  \includegraphics[width=.3\linewidth]{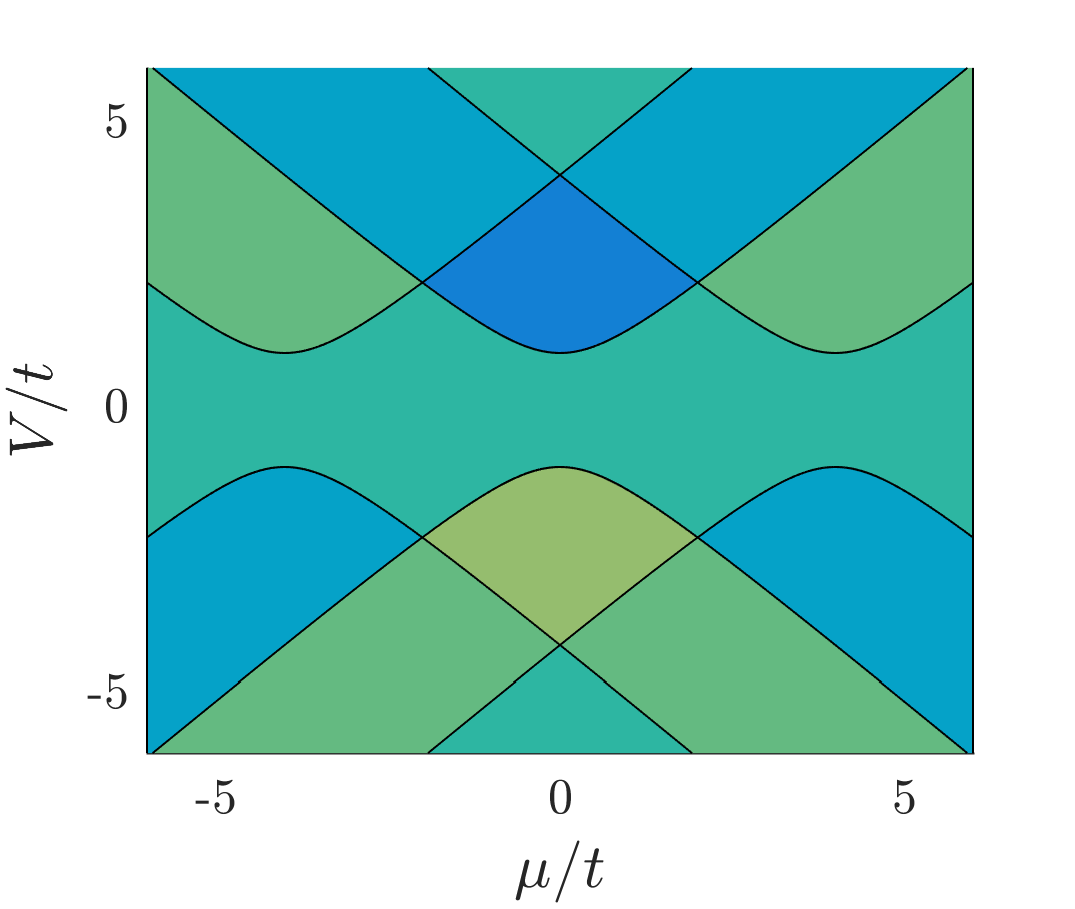}%
}\hfill
\subfloat[\label{fig:splusid_phase}]{%
  \includegraphics[width=.3\linewidth]{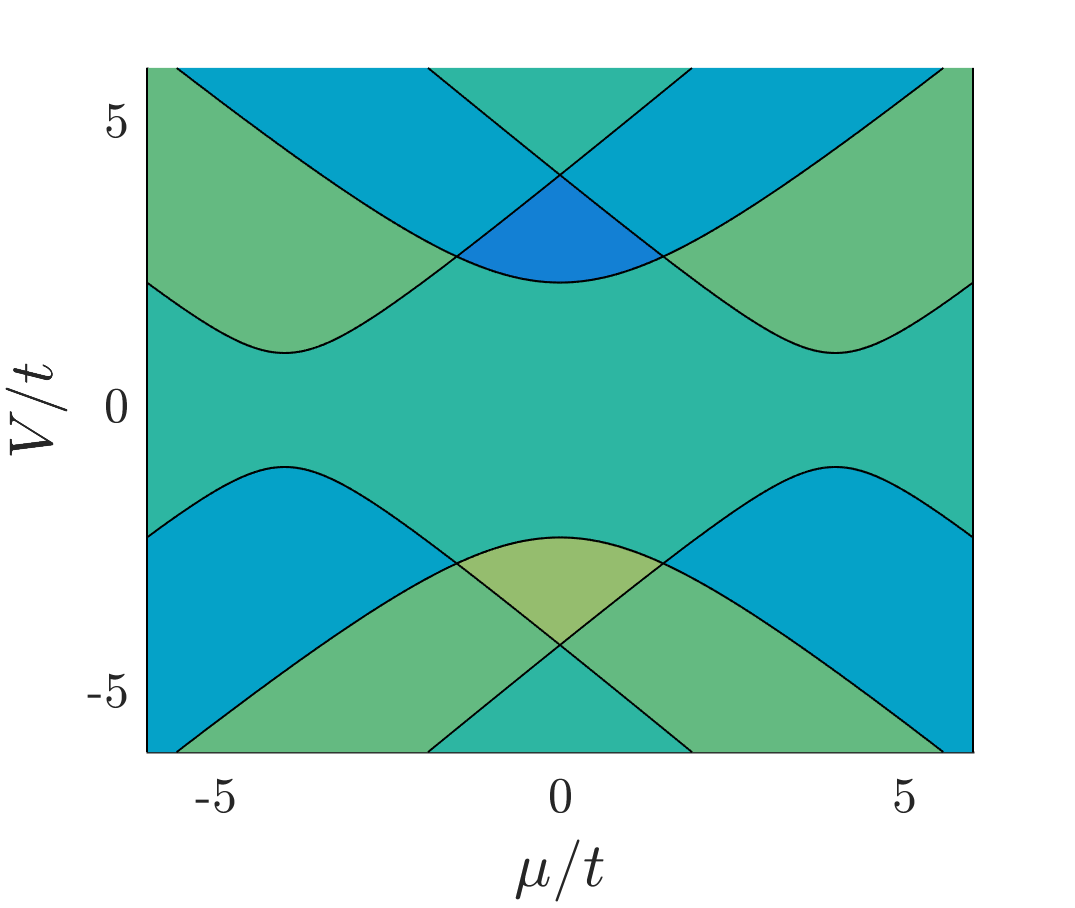}%
}\hfill
\subfloat[\label{fig:dplusid_phase}]{%
  \includegraphics[width=.3\linewidth]{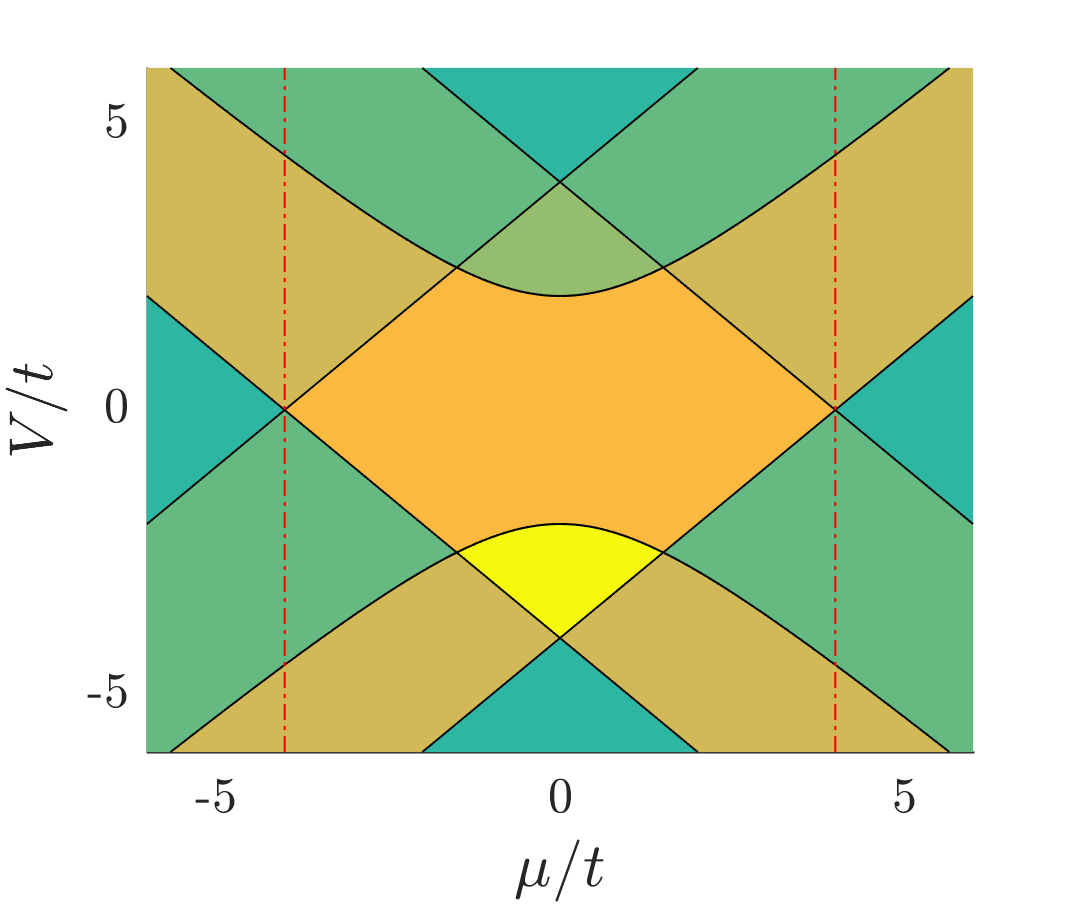}%
}\hfill
\subfloat{%
\raisebox{.775cm}
  {\input{colorbar.tikz}}
}
\caption{Total Chern number of the occupied bands, $\nu_{\mathrm{Ch}}=\nu_{\mathrm{Ch}}^-+\nu_{\mathrm{Ch}}^+$, following the definition of  Eq.\ \eqref{eq:chern}. (a) Phase diagram for $s+ig$ pairing. Parameters $\alpha=\Delta_{s}^0=\Delta_{g}^0=t$. Same diagram for $s$-wave and $s+id_{xy}$ cases. (b) Phase diagram for $s+id_{x^{2}-y^{2}}$ superconductor. Parameters $\alpha=\Delta_{s}^0=\Delta_{d_{x^{2}-y^{2}}}^0=t$. Parabola at the middle is shifted by $4\Delta_{d_{x^{2}-y^{2}}}^0$, in contrast to $s+ig$ case. (c) Phase diagram for $d+id$ superconductor. For $\left|\mu\right|<4t$, between the red dashed lines, we have two massive edge modes. This can be observed by computing the Chern number of just one of the bands. Parameters $\alpha=\Delta_{d_{x^{2}-y^{2}}}^0=\Delta_{d_{xy}}^0=t$.}
\label{fig:phase_diag}
\end{figure*}
We can also calculate the conditions for which the upper band gap, $E_+\left(\boldsymbol{k}\right)-E_-\left(\boldsymbol{k}\right)$, closes, i.e., $E_+\left(\boldsymbol{k}\right) = E_-\left(\boldsymbol{k}\right)$. We find
\begin{gather}
\epsilon\left(\boldsymbol{k}\right)=0,\label{eq: upper gap closing 1}\\ 
\left|\Delta\left(\boldsymbol{k}\right)\right|=0.\label{eq: upper gap closing 2}
\end{gather}

To characterize completely the phase diagram in these systems we need a topological invariant. The topological invariant associated with the different topological phases for 2D superconductors with no time-reversal symmetry is the Chern number \cite{Sato2016}. The bulk-edge correspondence relates the Chern number to the number of topological edge states at the boundary of the system \cite{TeoKane2010,Fukui2012}. We will numerically compute the Chern number by discretizing the Brillouin zone, using the expression \citep{Fukui2005,Resta2000}
\begin{equation}\label{eq:chern}
\nu_{\mathrm{Ch}}=\frac{1}{2\pi}\sum_{j}\text{Im}\left(\log\prod_{i}\langle\psi_{\boldsymbol{k}_{i}}|\psi_{\boldsymbol{k}_{i+1}}\rangle\right)_{j},
\end{equation}
where $|\psi_{\boldsymbol{k}_{i}}\rangle$ is an eigenvector of the Hamiltonian evaluated at $\boldsymbol{k}_{i}$, $j$ labels the different cells of the mesh that discretizes the Brillouin zone and $i$ runs over the four vertices of each cell. The expression in Eq.\ \eqref{eq:chern} is derived by integrating the Berry curvature over the 2D Brillouin zone. The Berry curvature can be approximated as $\frac{-1}{\delta A_j}\text{Im}\left(\log\prod_{i}\langle\psi_{\boldsymbol{k}_{i}}\psi_{\boldsymbol{k}_{i+1}}\rangle\right)_{j}$  at cell $j$, where $\delta A_j$ is the area of the cell and the value of Im is restricted to the principle branch of the logarithm. This definition yields a quantity which is manifestly gauge invariant.

We are interested in the singlet nodeless pairings compatible with point-group symmetry $D_{4h}$ of a tetragonal superconductor. They are given by the $s$-wave pairing and by four mixed pairings obtained by combining a real part from a 1D representation of the point group and an imaginary part from another 1D representation \cite{WengerPhysRevB.47.5977}. These 1D representations  (neglecting higher order terms) are
\begin{align}
&\Delta_s\left(\boldsymbol{k}\right) = \Delta_s^0,\\ 
&\Delta_g\left(\boldsymbol{k}\right)= \Delta_g^0\left(\sin2k_{x}\sin k_{y}-\sin2k_{y}\sin k_{x}\right),\\ 
&\Delta_{d_{xy}}\left(\boldsymbol{k}\right) = \Delta_{d_{xy}}^0\sin k_{x}\sin k_{y}, \\
&\Delta_{d_{x^2-y^2}}\left(\boldsymbol{k}\right) = \Delta_{d_{x^2-y^2}}^0\left(\cos k_{x}-\cos k_{y}\right).
\end{align}
Combining these representations we get the four possible mixed nodeless pairings
\begin{align}
&\Delta_{s}\left(\boldsymbol{k}\right)+i\Delta_{g}\left(\boldsymbol{k} \right),\hspace{3.6cm} \ \label{eq:sis}\\ 
&\Delta_{s}\left(\boldsymbol{k}\right)+i\Delta_{d_{xy}}\left(\boldsymbol{k}\right),\label{sidxy}\\
&\Delta_{s}\left(\boldsymbol{k}\right)+i\Delta_{d_{x^{2}-y^{2}}}\left(\boldsymbol{k}\right), \label{eq:sidxx}\\
&\Delta_{d_{x^{2}-y^{2}}}\left(\boldsymbol{k}\right)+i\Delta_{d_{xy}}\left(\boldsymbol{k}\right).\label{did}
\end{align}
In the following, they will be referred to as $s+ig$, $s+id_{xy}$, $s+id_{x^2-y^2}$ and $d+id$ respectively. All of them have various properties in common: they break time-reversal symmetry, they are nodeless and consequently they are characterized by the Chern number. Eq.\ \eqref{eq:sis} mixes $s$-wave and $g$-wave pairings and Eq.\ \eqref{did} is completely $d$-wave but not nodal. Eq.\ \eqref{sidxy} and Eq.\ \eqref{eq:sidxx} mix $d$-wave and $s$-wave pairings.

\section{Phase diagrams} \label{sec:phase_diag}
 This section is devoted to study the properties of the superconducting pairings shown in the previous section using phase diagrams. Each topological phase has a Chern number associated with it. Therefore, the following phase diagrams display the trivial and non-trivial phases which can be found upon varying the parameters of the Hamiltonian. By means of Eq.\ \eqref{eq:chern} we compute the topological phase diagrams  depicted in Fig.\ \ref{fig:phase_diag}  for the four possible pairings, where the total Chern number of the occupied bands, $\nu_{\mathrm{Ch}}=\nu_{\mathrm{Ch}}^-+\nu_{\mathrm{Ch}}^+$, is shown as a function of $\mu$ and $V$. $\nu_{\mathrm{Ch}}^\pm$ is computed by substituting in Eq.\ \eqref{eq:chern} the eigenvectors of the corresponding band. The pairings $s+ig$ and $s+id_{xy}$ share the same diagram. The Chern number value changes at points where the gap closes. Therefore we will solve Eqs.\ \eqref{eq: gap closing 1} and \eqref{eq: gap closing 2} to determine these gap-closing points. In particular, Eq.\ \eqref{eq: gap closing 2} is only satisfied when $\left|\boldsymbol{g}\left(\boldsymbol{k}\right)\right|=0$, since $\left|\Delta\left(\boldsymbol{k}\right)\right|=0$ is not possible for our nodeless pairings (except for the $d+id$ case at $\boldsymbol{k}=\left(0,0\right)$). The condition $\left|\boldsymbol{g}\left(\boldsymbol{k}\right)\right|=0$ implies that the momentum must be equal to $\boldsymbol{k}=\left(0,0\right),\left(0,\pi\right),\left(\pi,0\right),\left(\pi,\pi\right)$, which are the time-reversal invariant momenta. These four values for the momentum are then substituted in Eq.\  \eqref{eq: gap closing 1} yielding four gap-closing equations. Since $\boldsymbol{k}=\left(0,\pi\right)$ and $\boldsymbol{k}=\left(\pi,0\right)$ yield the same equation we effectively have three independent equations.
 
 In the following, we describe in detail the phase diagram for each nodeless pairing channel. The mixed pairing amplitudes $s+ig$ and $s+id_{xy}$ are analyzed within the same subsection since they are found to be topologically equivalent.
\subsection{$s+ig$ and $s+id_{xy}$} \label{sec:splusig}
We consider first the pairing $\Delta\left(\boldsymbol{k}\right)=\Delta_{s}^0+i\Delta_{g}^ 0\left(\sin2k_{x}\sin k_{y}-\sin2k_{y}\sin k_{x}\right)$. From Eqs.\ \eqref{eq: gap closing 1} and \eqref{eq: gap closing 2} we find the three equations where the gap closes, which are given by
\begin{align}
&V^{2}=\left(4t+\mu\right)^{2}+\left(\Delta_{s}^0\right)^{2}, \label{eq:s+ig closing 1}\\
&V^{2}=\mu^{2}+\left(\Delta_{s}^0\right)^{2}, \label{eq:s+ig closing 2}\\
&V^{2}=\left(4t-\mu\right)^{2}+\left(\Delta_{s}^0\right)^{2} \label{eq:s+ig closing 3}.
\end{align}
The upper gap between the upper and lower bands, $E_+\left(\boldsymbol{k}\right)-E_-\left(\boldsymbol{k}\right)$, does not close since the condition $\left|\Delta\left(\boldsymbol{k}\right)\right|=0$ given in Eq.\ \eqref{eq: upper gap closing 2} cannot be satisfied. $\Delta_g^0$ does not appear in Eqs.\ \eqref{eq:s+ig closing 1}-\eqref{eq:s+ig closing 3} because $\Delta_{g}\left(\boldsymbol{k} \right)$ vanishes at  the time-reversal invariant momenta. A continuous deformation of the Hamiltonian taking $\Delta_{g}^0\rightarrow0$ does not close the gap. Thus, the $\Delta_{s}\left(\boldsymbol{k}\right)+i\Delta_{g}\left(\boldsymbol{k} \right)$
superconductor and a conventional $s$-wave superconductor are topologically
equivalent. 

For the $s+id_{xy}$ case, $\Delta\left(\boldsymbol{k}\right)=\Delta_{s}^0+i\Delta_{d_{xy}}^0\sin k_{x}\sin k_{y}$, we obtain the same gap closing equations that we had for the $s+ig$
case, since $\Delta_{d_{xy}}\left(\boldsymbol{k}\right)$ vanishes
at $\boldsymbol{k}=\left(0,0\right),\left(0,\pi\right),\left(\pi,0\right),\left(\pi,\pi\right)$.
This means we can also take $\Delta_{d_{xy}}^0\rightarrow0$ without closing
the gap. In the phase diagram of Fig.\ \ref{fig:splusis_phase} we depict the different transition points given by Eqs.\ \eqref{eq:s+ig closing 1}-\eqref{eq:s+ig closing 3}. The Chern number takes values between -2 and 2, which, by means of the bulk-edge correspondence, implies that the system can host up to two edge states.

\subsection{$s+id_{x^{2}-y^{2}}$}
Considering the $s+id_{x^{2}-y^{2}}$ pairing, $\Delta\left(\boldsymbol{k}\right)=\Delta_{s}^0+i\Delta_{d_{x^{2}-y^{2}}}^0\left(\cos k_{x}-\cos k_{y}\right)$, we can compute the gap closing points as we did for $s+ig$. We obtain the same equations as we had in Sec. \ref{sec:splusig} (Eqs.\ \eqref{eq:s+ig closing 1} and \eqref{eq:s+ig closing 3}) but instead of Eq.\ \eqref{eq:s+ig closing 2} we now obtain
\begin{equation}
V^{2}=\mu^{2}+\left(\Delta_{s}^0\right)^{2}+4\left(\Delta_{d_{x^{2}-y^{2}}}^0\right)^{2}. \label{eq:s+idxy closing}
\end{equation}
In contrast to $s+ig$ or $s+id_{xy}$, where neither $\Delta_{g}^0$ nor $\Delta_{d_{xy}}^0$ played any role, the pairing amplitude $\Delta_{d_{x^{2}-y^{2}}}^0$ appears in Eq.\ \eqref{eq:s+idxy closing}.  The upper gap does not close. Notably, the Chern number takes the same values as $s+ig$, see phase diagram in Fig.\ \ref{fig:splusid_phase}. However, now we have that the middle parabola given by  Eq.\ \eqref{eq:s+idxy closing} is shifted because of $\Delta_{d_{x^{2-y^{2}}}}^0$, in contrast to $s+ig$ and $s+id_{xy}$.

\subsection{$d+id$}
For $d+id$ pairing, $\Delta\left(\boldsymbol{k}\right)=\Delta_{d_{x^{2}-y^{2}}}^0\left(\cos k_{x}-\cos k_{y}\right)+i\Delta_{d_{xy}}^0\sin k_{x}\sin k_{y}$, substituting in Eqs.\ \eqref{eq: gap closing 1} and \eqref{eq: gap closing 2} one finds the gap-closing conditions
\begin{align}
&V^{2}=\left(4t+\mu\right)^{2}, \\
&V^{2} = \mu^{2}+4\left(\Delta_{d_{x^{2}-y^{2}}}^0\right)^{2},\\
&V^{2} = \left(4t-\mu\right)^{2}.
\end{align}
Remarkably in this case, the upper gap between the two particle bands closes, unlike what happens for the other pairing channels. The condition $\left|\Delta\left(\boldsymbol{k}\right)\right|=0$ of Eq.\ \eqref{eq: upper gap closing 2} implies $\boldsymbol{k}=\left(0,0\right),\left(\pi,\pi\right)$. Substituting
these values into Eq.\ \eqref{eq: upper gap closing 1}, $\epsilon\left(\boldsymbol{k}\right)=0$, we get
$\mu=\pm4t$. Therefore the Chern number of the upper band takes non-zero
values for $-4t<\mu<4t$, in particular we have $\nu_\mathrm{Ch}^+=2$ (the Chern number of the upper band for other
pairings was zero). The Chern number of the
occupied bands is plotted in Fig.\ \ref{fig:dplusid_phase}, taking values between 0 and 6, in contrast to what we found in previous cases where we had values between -2 and 2.

\begin{figure}
\subfloat[\label{fig:splusis1}]{%
  \includegraphics[width=.49\linewidth]{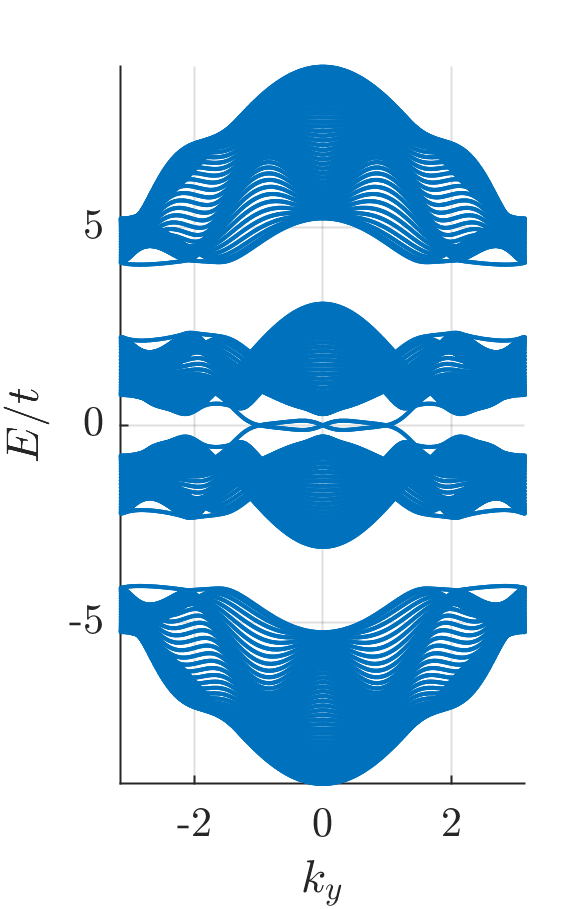}%
}\hfill
\subfloat[\label{fig:splusid2}]{%
  \includegraphics[width=.49\linewidth]{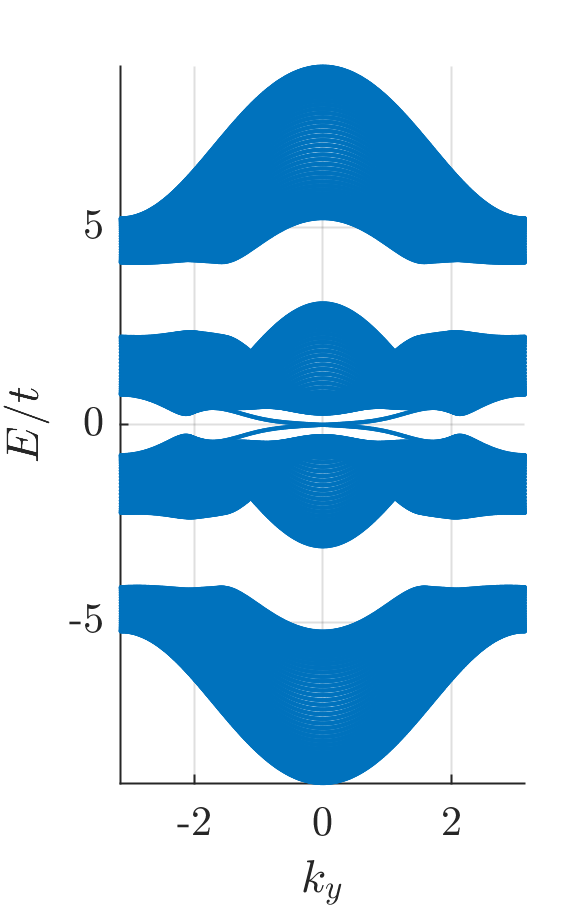}%
}
\caption{$s+ig$ spectrum on a cylindrical geometry. Chern number is 1. The edge states in (a) cross zero energy three times. By taking $\Delta_g^0$ from $2t$ to $0.5t$, which is a smooth deformation, we see how the zero-energy modes at $k_y\neq0$ disappear in (b). Parameters: $\mu=2t$, $V=3t$, $\alpha=\Delta_{s}^0=t$, lattice sites $N_x=50$. } \label{fig:splusis_spectrum}
\end{figure}

\section{Edge states and spectra} \label{sec:edge_states}

In this section, we investigate the connection between the Chern number and the physics of the edge states, when considering our previous system with open boundary conditions. Therefore, we place our model Hamiltonian on a cylindrical geometry with open boundary conditions in the $x$-direction and periodic boundary conditions in the $y$-direction. In this way, $k_y$ remains a good quantum number and we can observe the edge states appearing at the boundaries of the cylinder. In Sec. \ref{sec:Chern_zero_energy} we study the connection between the Chern number, the edge states and the zero-energy modes. In Sec. \ref{sec:edge_dplusid} we show particularly interesting features of the $d+id$ spectrum.

\subsection{Chern number and zero-energy modes} \label{sec:Chern_zero_energy}
The bulk-boundary correspondence  establishes a relation between the bulk Chern number and the edge states at the boundary of the system. This is given explicitly by the equation \citep{Fukui2012,RevModPhysRyu}
\begin{equation}\label{eq:spectral_flow}
\nu_{\mathrm{Ch}} = N_+ - N_-,
\end{equation}
where $N_+$ ($N_-$) is the number of forward (backward) propagating edge states. The above formula relates the Chern number with the spectral flow, i.e., the difference between the number of edge states connecting the negative energy band with the positive energy band (forward propagating $N_+$) and the number of edge states going the other way around (backward propagating $N_-$). In general, we will expect for our model that the number of edge states at one boundary equals the Chern number.

One may be tempted to equate the number of edge states to the number of zero-energy modes, since the edge states cross the zero-energy level on their way across the gap. Although this is usually true, we have found some exceptions in the $s+ig$ and $d+id$ cases where one of the edge states crosses the zero-energy level more than once, producing three zero-energy modes. These three zero-energy modes cannot be expected to be topologically robust, since a smooth deformation of the Hamiltonian, such as the introduction of weak disorder, can reduce the number of crossings to one. Thus, only one of the zero-modes can be qualified as topologically robust. A particular example of this can be seen in Fig.\ \ref{fig:splusis_spectrum} for the case of $s+ig$, where only the crossing at $k_y=0$ is topologically robust. By tuning $\Delta_g^0$ the number of zero modes is reduced to one.

As a consequence of the bulk-boundary correspondence, the edge states are topologically protected and are robust under weak static disorder. If disorder is weak and respects particle-hole symmetry, the gap does not close and the topological states propagating at the edge of the system remain, since Eq.\ \eqref{eq:spectral_flow} still holds. This has been tested by introducing weak disorder perturbations to the parameters of the Hamiltonian and diagonalising numerically to obtain the edge states. 
\subsection{Edge states in $d+id$} \label{sec:edge_dplusid}

\begin{figure}
\subfloat[\label{fig:dplusid1}]{%
  \includegraphics[width=.49\linewidth]{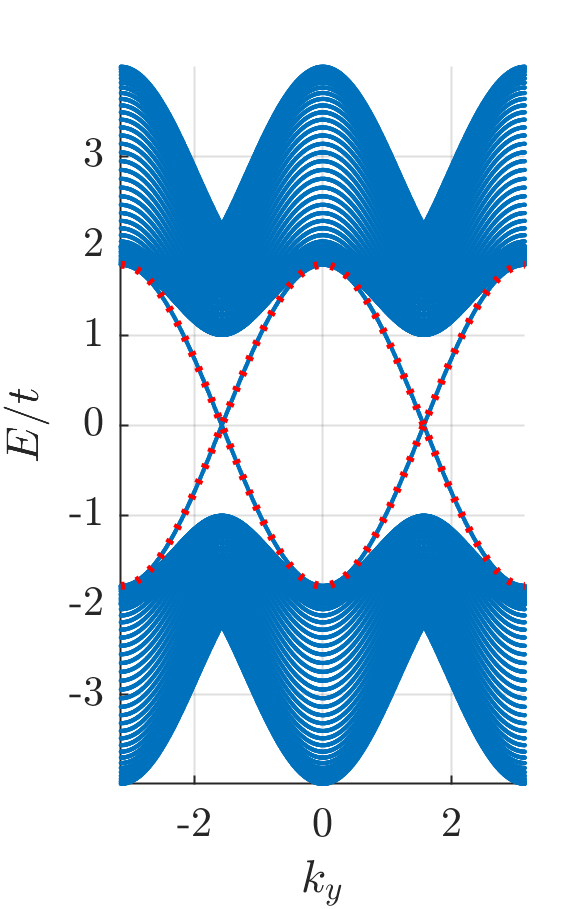}%
}\hfill
\subfloat[\label{fig:dplusid2}]{%
  \includegraphics[width=.49\linewidth]{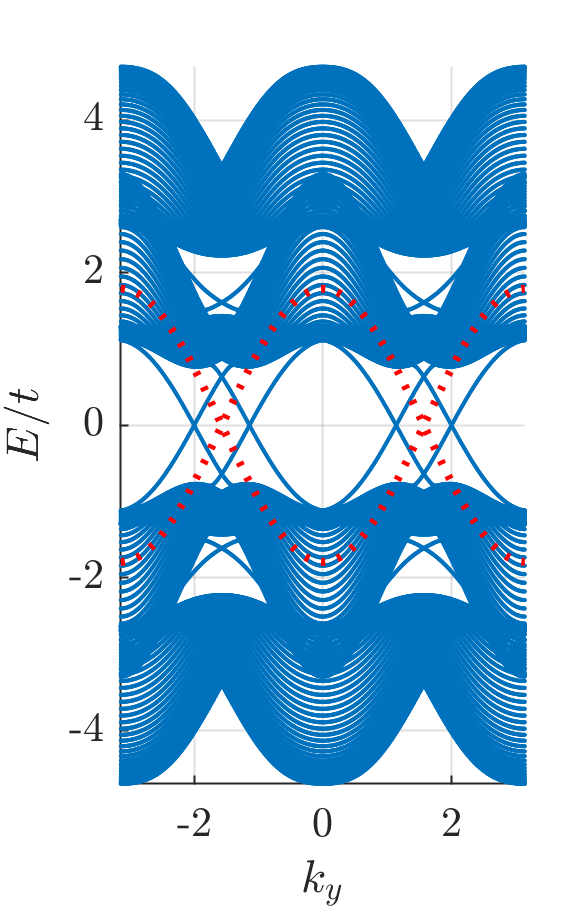}%
}
\caption{$d+id$ spectra on a cylindrical geometry. Chern number is 4, which corresponds to the four edge states that cross zero energy at each boundary of the cylinder. (a) No SOC nor Zeeman field, bands are degenerate. The dotted line are the edge states given by Eq.\ \eqref{eq:edge_dispersion}. (b) SOC and Zeeman field are non-zero ($\alpha =t$, $V=0.7t$). The bands and the edge modes move with respect to the original ones, still depicted in red. Note the two massive edge modes between the upper and the lower band. Parameters: $\mu=0$, $\Delta_{d_{x^{2}-y^{2}}}^0=\Delta_{d_{xy}}^0=t$, $N_x=50$. \label{fig:dplusid_spectrum}}
\end{figure}

In this section, we study the $d+id$ pairing channel in detail. One noteworthy characteristic  of the $d+id$ spectrum is the presence of zero-energy modes away of the time-reversal invariant momenta $k_y = 0,\pi$, where they appear for the other pairings considered. This new phenomena occur for regions of the topological phase diagram with $\nu_{\mathrm{Ch}}=3,4,6$. We will now explain why this happens for $\nu_{\mathrm{Ch}}=4$ and small Zeeman field (central part of Fig.\ \ref{fig:dplusid_phase}), by applying some ideas of bulk-boundary correspondence \cite{Mong2011,Hatsugai2002}. In this region we have four edge states with four zero-energy modes at $k_y\neq0,\pi$, even in the absence of Zeeman field or SOC.

We start by considering the $d+id$ pairing without SOC nor Zeeman field. Hamiltonian \eqref{eq:H_sum} can be separated in two independent subsystems, ${\cal H}={\cal H}_\uparrow+{\cal H}_\downarrow$. Thus we have
\begin{equation}\label{eq:H_sum_2}
{\cal H}_\uparrow=\frac{1}{2}\sum_{\boldsymbol{k}}\left(c_{\boldsymbol{k},\uparrow}^{\dagger},c_{-\boldsymbol{k},\downarrow}\right)H_\uparrow\left(\boldsymbol{k}\right)\left(\begin{array}{c}
c_{\boldsymbol{k},\uparrow}\\
c_{-\boldsymbol{k},\downarrow}^{\dagger}
\end{array}\right), 
\end{equation}
with
\begin{equation} \label{eq:H_matrix_2}
H_\uparrow\left(\boldsymbol{k}\right)=
\left(\begin{array}{cc}
\epsilon\left(\boldsymbol{k}\right) & \Delta\left(\boldsymbol{k}\right)\\
\Delta^{*}\left(\boldsymbol{k}\right) & -\epsilon\left(\boldsymbol{k}\right)
\end{array}\right).
\end{equation}
${\cal H}_\downarrow$ can be obtained by turning the spins upside down and yields the same spectrum as ${\cal H}_\uparrow$, i.e., the bands of ${\cal H}$ are now degenerate. We will now drop the $\uparrow$ and $\downarrow$ symbols for simplicity. Eq.\ \eqref{eq:H_matrix_2} can be recast into the form $H=\boldsymbol{h}\cdot\boldsymbol{\sigma}$, where $\boldsymbol{h}=\left(\Delta_{d_{x^{2}-y^{2}}}\left(\boldsymbol{k}\right),-\Delta_{d_{xy}}\left(\boldsymbol{k}\right),\epsilon\left(\boldsymbol{k}\right)\right)$ and $\boldsymbol{\sigma}$ are the Pauli matrices. Since we are interested in the spectrum on a cylindrical geometry with periodic boundary conditions in $y$-direction and open boundary conditions in $x$-direction, we will express $H$ as a family of one-dimensional Hamiltonians labeled by $k_y$, i.e., $H_{k_y}(k_x)$. Thereby, $\boldsymbol{h}_{k_y}(k_x)$ defines a map from the 1D Brillouin zone to a loop in $\mathbb{R}^3$. Note that the origin of $\mathbb{R}^3$ corresponds to the closing of the gap. It is straightforward to see that the loop defined by $\boldsymbol{h}_{k_y}(k_x)$ is an ellipse contained in a plane by reexpressing $\boldsymbol{h}_{k_y}(k_x)$ as
\begin{equation}
\boldsymbol{h}_{k_y}(k_x) = \boldsymbol{b}_0+ \boldsymbol{b}_1 \cos k_x+ \boldsymbol{b}_2 \sin k_x,
\end{equation}
where $\boldsymbol{b}_1 \cos k_x$ and $\boldsymbol{b}_2 \sin k_x$ span the ellipse on a plane with normal vector $\boldsymbol{n}_\perp = \boldsymbol{b}_1 \times \boldsymbol{b}_2 / \left| \boldsymbol{b}_1 \times \boldsymbol{b}_2 \right|$ and $\boldsymbol{b}_0$ shifts the center of the ellipse from the origin. These vectors are given by
\begin{align}
&\boldsymbol{b}_0 = \left(-\Delta^0_{d_{x^2-y^2}}\cos k_y,0,-2t\cos k_y-\mu\right), \label{eq:b0}\\
&\boldsymbol{b}_1 \cos k_x=\left(\Delta^0_{d_{x^2-y^2}},0,-2t\right) \cos k_x, \label{eq:b1}\\
&\boldsymbol{b}_2 \sin k_x=\left(0,-\Delta^0_{d_{xy}}\sin k_y,0\right) \sin k_x. \label{eq:b2}
\end{align}
The vector $\boldsymbol{b}_0$ can be devided into two contributions. The first one shifts the ellipse from the origin in direction parallel to the plane containing the loop, $\boldsymbol{b}_0^\parallel$. The other contribution shifts it perpendicularly, $\boldsymbol{b}_0^\perp$.

It can be proven that if the loop defined by $\boldsymbol{h}_{k_y}(k_x)$ contains the origin, the Hamiltonian \eqref{eq:H_sum_2} has a zero-energy mode when placed on a chain with open boundary conditions. To show this, we first move the loop to the $xy$-plane in $\mathbb{R}^3$ by making a simple rotation. This is equivalent to performing a SU(2) transformation to our Nambu basis, $\boldsymbol{d}_{k_x}=U\left(c_{k_x,\uparrow},c^\dagger_{-k_x,\downarrow}\right)^T$. Next, we can smoothly deform the loop into a circle without crossing the origin (since this would close the gap). After this smooth deformation, we obtain $\boldsymbol{h}_{k_y}(k_x)=\left(\cos k_x, \sin k_x, 0\right)$. Fourier transforming this Hamiltonian back to real space we find that for an open chain
\begin{equation}
{\cal H}'=\sum_{n=1}^{N_x-1} d^\dagger_{n,\uparrow} d^\dagger_{n+1,\downarrow} + \mathrm{H.c.},
\end{equation}
where we can immediately see that $d^\dagger_{1,\downarrow}$, $d_{1,\downarrow}$, $d^\dagger_{N_x,\uparrow}$, and $d_{N_x,\uparrow}$ do not appear, so that we have zero-energy modes localized at the edges. 

One can also prove \cite{Mong2011} that even if the loop does not contain the origin, we have edge states with non-zero energy if the projection of the loop in the perpendicular direction, $\boldsymbol{n}_\perp$, does contain it. The energy of these edge states is given by the perpendicular distance to the origin $E=\pm | \boldsymbol{b}_0\cdot \boldsymbol{n}^\perp |$. For the $d+id$ case we have
\begin{equation}\label{eq:edge_dispersion}
E\left(k_y\right) = \pm  \frac{4t \cos k_y +\mu }{\sqrt{4t^2+\left(\Delta^0_{d_{x^2-y^2}}\right)^2}}\Delta^0_{d_{x^2-y^2}}.
\end{equation}
This implies the existence of zero-energy modes at $\cos k_y =\pm\frac{\mu}{4t}$, which shows that we have solutions at $k_y\neq 0,\pi$.

The results we have obtained are valid for $d+id$ paring without SOC nor Zeeman field. The spectrum for this particular setting can be seen in Fig.\ \ref{fig:dplusid1}. When a small Zeeman field is turned on, ${\cal H}_\uparrow$ and ${\cal H}_\downarrow$ no longer have the same spectrum and the degeneracy breaks down. One band moves up in energy and the other one moves down doubling the number of Dirac cones. This can be accounted for in Eq.\ \eqref{eq:edge_dispersion} by adding a term $\pm V$. The effect of SOC is rather complicated but it can be roughly described as a horizontal displacement of the bands with respect to one another, that also doubles the number of Dirac cones. Combining the Zeeman field and the SOC, the zero-energy states are placed further apart from each other in momentum space. The final result is four zero-energy modes away from $k_y=0,\pi$, see Fig.\ \ref{fig:dplusid2}. Remarkably, even with no Zeeman field or SOC the Chern number is non-zero and there are are topologically protected edge states. Note that in this case, weak disorder can slightly displace the momentum $k_y$ at which the edge states cross the zero-energy level. However, these edge states cannot be smoothly connected to the bulk, since they are topologically robust. Hence, these edge states always cross the zero-energy level and connect two bulk bands, despite the crossing point might be slightly shifted due to disorder, as described in Sec. \ref{sec:Chern_zero_energy}.

Another interesting feature of the $d+id$ spectrum appears when we have a non-zero Chern number for the upper band. In the region  $|\mu|<4t$ we have $\nu_\mathrm{Ch}^+=2$, and we can observe two massive edge modes in addition to the zero-energy modes that may appear, as can be seen in Fig.\ \ref{fig:dplusid2}. This effect is something unique of the $d+id$ pairing symmetry.

\section{Conclusions} \label{sec:conclusions}
We have studied four 2D tetragonal superconductors with SOC and Zeeman field. The four possible mixed nodeless pairings compatible with the point-group symmetry of tetragonal superconductors have been considered. For these systems we are able to obtain the phase diagrams by numerically computing the Chern number and calculating the gap closing conditions. The phase diagrams for $s+ig$ and $s+id_{xy}$ are found to be equivalent to the  $s$-wave case, since we can continuously deform the Hamiltonian without closing the gap. The $d+id$ pairing channel shows rich behavior, with the Chern number taking values up to 6.

Computing the energy spectra for the different pairing amplitudes, we have found that for certain cases the number of zero-energy modes is higher than the Chern number. These extra zero-energy modes are shown to be topologically unprotected and disappear in the presence of disorder. For the $d+id$ case we have explained why zero-modes appearing at $k\neq0,\pi$ are possible, in contrast to what occurs for other cases, where we have $k=0,\pi$. Finally, we have also seen how massive edge modes appear associated to the non-trivial topology of the upper band. This is proven by calculating the Chern number of the upper band.

Potential applications of anysotropic $d+id$ superconducting pairings have been proposed in materials such as water-intercalated sodium cobaltates, bilayer silicene, epitaxial bilayer films of bismuth and nickel or FeAs-based superconductors \cite{PhysRevLett.111.097001, PhysRevLett.111.066804,doi:10.1063/1.4961462,PhysRevB.94.064519, PhysRevLett.100.217002,Gong2017}.  Our results could also be applied to a broaden context: the new physics emerging in multicomponent superconducting systems, with a multicomponent superconducting order parameter as a consequence of particular pairing symmetry \cite{0953-2048-28-6-060201}. 

Interestingly, the ultra-highly doped monolayer graphene \cite{natphys8} was proposed to support $d+id$ superconductivity originated from repulsive electron-electron interactions. This begs the question of whether a similar analysis to the one performed in this paper, but for the corresponding dihedral symmetry group, could shed some light over the recently discovered superconducting twisted bilayer graphene \cite{jarillo,Fu_conclusions}.
\begin{acknowledgments} 
We acknowledge financial support from the Spanish MINECO grants FIS2012-33152, FIS2015-67411, and the CAM research consortium QUITEMAD+, Grant No. S2013/ICE-2801. The research of M.A.M.-D. has been supported in part by the U.S. Army Research Office through Grant No. W911N F-14-1-0103.
O.V. thanks Hiroki Isobe for interesting discussions, Fundaci\'on Ram\'on Areces and RCC Harvard. S.V. thanks FPU MECD Grant. 
\end{acknowledgments}

\bibliography{citations.bib}

\end{document}

%% file: colorbar.tikz
\def\h{3.4}
\def\w{.5}
\def\ntick{8}
\def\nticks{6}
\begin{tikzpicture}

\path[fill=mtwo] (0,0) -- (\w,0) -- (\w,{\h/(\ntick+1)})-- (0,{\h/(\ntick+1)})  -- cycle;
\path[fill=mone] (0,{\h/(\ntick+1)}) -- (\w,{\h/(\ntick+1)}) -- (\w,{2*\h/(\ntick+1)})-- (0,{2*\h/(\ntick+1)})  -- cycle;
\path[fill=zero] (0,{2*\h/(\ntick+1)}) -- (\w,{2*\h/(\ntick+1)}) -- (\w,{3*\h/(\ntick+1)})-- (0,{3*\h/(\ntick+1)})  -- cycle;
\path[fill=one] (0,{3*\h/(\ntick+1)}) -- (\w,{3*\h/(\ntick+1)}) -- (\w,{4*\h/(\ntick+1)})-- (0,{4*\h/(\ntick+1)})  -- cycle;
\path[fill=two] (0,{4*\h/(\ntick+1)}) -- (\w,{4*\h/(\ntick+1)}) -- (\w,{5*\h/(\ntick+1)})-- (0,{5*\h/(\ntick+1)})  -- cycle;
\path[fill=three] (0,{5*\h/(\ntick+1)}) -- (\w,{5*\h/(\ntick+1)}) -- (\w,{6*\h/(\ntick+1)})-- (0,{6*\h/(\ntick+1)})  -- cycle;
\path[fill=four] (0,{6*\h/(\ntick+1)}) -- (\w,{6*\h/(\ntick+1)}) -- (\w,{7*\h/(\ntick+1)})-- (0,{7*\h/(\ntick+1)})  -- cycle;
\path[fill=five] (0,{7*\h/(\ntick+1)}) -- (\w,{7*\h/(\ntick+1)}) -- (\w,{8*\h/(\ntick+1)})-- (0,{8*\h/(\ntick+1)})  -- cycle;
\path[fill=six] (0,{8*\h/(\ntick+1)}) -- (\w,{8*\h/(\ntick+1)}) -- (\w,{9*\h/(\ntick+1)})-- (0,{9*\h/(\ntick+1)})  -- cycle;

\foreach \i in {1,...,\ntick}
{
	\draw (\w,{\i*\h/(\ntick+1)}) -- (\w-\w/6,{\i*\h/(\ntick+1)}) ;
}

\foreach \i in {-2,...,\nticks}
{
	\node[right,font=\footnotesize] at (\w,{(\i+3-1)*\h/(\ntick+1)+\h/(\ntick+1)/2}) {\i};
}

\draw (0,0) -- (\w,0) -- (\w,\h) -- (0,\h) -- (0,0);

\end{tikzpicture}

%% file: nodeless.bbl
\begin{thebibliography}{55}%
\makeatletter
\providecommand \@ifxundefined [1]{%
 \@ifx{#1\undefined}
}%
\providecommand \@ifnum [1]{%
 \ifnum #1\expandafter \@firstoftwo
 \else \expandafter \@secondoftwo
 \fi
}%
\providecommand \@ifx [1]{%
 \ifx #1\expandafter \@firstoftwo
 \else \expandafter \@secondoftwo
 \fi
}%
\providecommand \natexlab [1]{#1}%
\providecommand \enquote  [1]{``#1''}%
\providecommand \bibnamefont  [1]{#1}%
\providecommand \bibfnamefont [1]{#1}%
\providecommand \citenamefont [1]{#1}%
\providecommand \href@noop [0]{\@secondoftwo}%
\providecommand \href [0]{\begingroup \@sanitize@url \@href}%
\providecommand \@href[1]{\@@startlink{#1}\@@href}%
\providecommand \@@href[1]{\endgroup#1\@@endlink}%
\providecommand \@sanitize@url [0]{\catcode `\\12\catcode `\$12\catcode
  `\&12\catcode `\#12\catcode `\^12\catcode `\_12\catcode `\%12\relax}%
\providecommand \@@startlink[1]{}%
\providecommand \@@endlink[0]{}%
\providecommand \url  [0]{\begingroup\@sanitize@url \@url }%
\providecommand \@url [1]{\endgroup\@href {#1}{\urlprefix }}%
\providecommand \urlprefix  [0]{URL }%
\providecommand \Eprint [0]{\href }%
\providecommand \doibase [0]{http://dx.doi.org/}%
\providecommand \selectlanguage [0]{\@gobble}%
\providecommand \bibinfo  [0]{\@secondoftwo}%
\providecommand \bibfield  [0]{\@secondoftwo}%
\providecommand \translation [1]{[#1]}%
\providecommand \BibitemOpen [0]{}%
\providecommand \bibitemStop [0]{}%
\providecommand \bibitemNoStop [0]{.\EOS\space}%
\providecommand \EOS [0]{\spacefactor3000\relax}%
\providecommand \BibitemShut  [1]{\csname bibitem#1\endcsname}%
\let\auto@bib@innerbib\@empty
\bibitem [{\citenamefont {Thouless}\ \emph {et~al.}(1982)\citenamefont
  {Thouless}, \citenamefont {Kohmoto}, \citenamefont {Nightingale},\ and\
  \citenamefont {den Nijs}}]{PhysRevLett.49.405}%
  \BibitemOpen
  \bibfield  {author} {\bibinfo {author} {\bibfnamefont {D.~J.}\ \bibnamefont
  {Thouless}}, \bibinfo {author} {\bibfnamefont {M.}~\bibnamefont {Kohmoto}},
  \bibinfo {author} {\bibfnamefont {M.~P.}\ \bibnamefont {Nightingale}}, \ and\
  \bibinfo {author} {\bibfnamefont {M.}~\bibnamefont {den Nijs}},\ }\href
  {\doibase 10.1103/PhysRevLett.49.405} {\bibfield  {journal} {\bibinfo
  {journal} {Phys. Rev. Lett.}\ }\textbf {\bibinfo {volume} {49}},\ \bibinfo
  {pages} {405} (\bibinfo {year} {1982})}\BibitemShut {NoStop}%
\bibitem [{\citenamefont {Haldane}(1983)}]{HALDANE1983464}%
  \BibitemOpen
  \bibfield  {author} {\bibinfo {author} {\bibfnamefont {F.}~\bibnamefont
  {Haldane}},\ }\href {\doibase 10.1016/0375-9601(83)90631-X} {\bibfield
  {journal} {\bibinfo  {journal} {Physics Letters A}\ }\textbf {\bibinfo
  {volume} {93}},\ \bibinfo {pages} {464 } (\bibinfo {year}
  {1983})}\BibitemShut {NoStop}%
\bibitem [{\citenamefont {Wen}(1992)}]{doi:10.1142/S0217979292000840}%
  \BibitemOpen
  \bibfield  {author} {\bibinfo {author} {\bibfnamefont {X.-G.}\ \bibnamefont
  {Wen}},\ }\href {\doibase 10.1142/S0217979292000840} {\bibfield  {journal}
  {\bibinfo  {journal} {International Journal of Modern Physics B}\ }\textbf
  {\bibinfo {volume} {06}},\ \bibinfo {pages} {1711} (\bibinfo {year}
  {1992})}\BibitemShut {NoStop}%
\bibitem [{\citenamefont {Gonzalez}\ \emph {et~al.}(1995)\citenamefont
  {Gonzalez}, \citenamefont {Martin-Delgado}, \citenamefont {Sierra},\ and\
  \citenamefont {Vozmediano}}]{book_martin-delgado}%
  \BibitemOpen
  \bibfield  {author} {\bibinfo {author} {\bibfnamefont {J.}~\bibnamefont
  {Gonzalez}}, \bibinfo {author} {\bibfnamefont {M.}~\bibnamefont
  {Martin-Delgado}}, \bibinfo {author} {\bibfnamefont {G.}~\bibnamefont
  {Sierra}}, \ and\ \bibinfo {author} {\bibfnamefont {A.~H.}\ \bibnamefont
  {Vozmediano}},\ }\href {\doibase 10.1007/978-3-540-47678-8} {\emph {\bibinfo
  {title} {Quantum Electron Liquids and High-$T_c$ superconductivity}}}\
  (\bibinfo  {publisher} {Spinger-Berlin},\ \bibinfo {year} {1995})\BibitemShut
  {NoStop}%
\bibitem [{\citenamefont {Nayak}\ \emph {et~al.}(2008)\citenamefont {Nayak},
  \citenamefont {Simon}, \citenamefont {Stern}, \citenamefont {Freedman},\ and\
  \citenamefont {Das~Sarma}}]{RevModPhys.80.1083}%
  \BibitemOpen
  \bibfield  {author} {\bibinfo {author} {\bibfnamefont {C.}~\bibnamefont
  {Nayak}}, \bibinfo {author} {\bibfnamefont {S.~H.}\ \bibnamefont {Simon}},
  \bibinfo {author} {\bibfnamefont {A.}~\bibnamefont {Stern}}, \bibinfo
  {author} {\bibfnamefont {M.}~\bibnamefont {Freedman}}, \ and\ \bibinfo
  {author} {\bibfnamefont {S.}~\bibnamefont {Das~Sarma}},\ }\href {\doibase
  10.1103/RevModPhys.80.1083} {\bibfield  {journal} {\bibinfo  {journal} {Rev.
  Mod. Phys.}\ }\textbf {\bibinfo {volume} {80}},\ \bibinfo {pages} {1083}
  (\bibinfo {year} {2008})}\BibitemShut {NoStop}%
\bibitem [{\citenamefont {Alicea}(2012)}]{RepAlicea}%
  \BibitemOpen
  \bibfield  {author} {\bibinfo {author} {\bibfnamefont {J.}~\bibnamefont
  {Alicea}},\ }\href {http://stacks.iop.org/0034-4885/75/i=7/a=076501}
  {\bibfield  {journal} {\bibinfo  {journal} {Reports on Progress in Physics}\
  }\textbf {\bibinfo {volume} {75}},\ \bibinfo {pages} {076501} (\bibinfo
  {year} {2012})}\BibitemShut {NoStop}%
\bibitem [{\citenamefont {Litinski}\ \emph {et~al.}(2017)\citenamefont
  {Litinski}, \citenamefont {Kesselring}, \citenamefont {Eisert},\ and\
  \citenamefont {von Oppen}}]{PhysRevX.7.031048}%
  \BibitemOpen
  \bibfield  {author} {\bibinfo {author} {\bibfnamefont {D.}~\bibnamefont
  {Litinski}}, \bibinfo {author} {\bibfnamefont {M.~S.}\ \bibnamefont
  {Kesselring}}, \bibinfo {author} {\bibfnamefont {J.}~\bibnamefont {Eisert}},
  \ and\ \bibinfo {author} {\bibfnamefont {F.}~\bibnamefont {von Oppen}},\
  }\href {\doibase 10.1103/PhysRevX.7.031048} {\bibfield  {journal} {\bibinfo
  {journal} {Phys. Rev. X}\ }\textbf {\bibinfo {volume} {7}},\ \bibinfo {pages}
  {031048} (\bibinfo {year} {2017})}\BibitemShut {NoStop}%
\bibitem [{\citenamefont {Volovik}(1988)}]{Volovik1988}%
  \BibitemOpen
  \bibfield  {author} {\bibinfo {author} {\bibfnamefont {G.}~\bibnamefont
  {Volovik}},\ }\href@noop {} {\bibfield  {journal} {\bibinfo  {journal} {Sov.
  Phys. JETP}\ }\textbf {\bibinfo {volume} {67}},\ \bibinfo {pages} {1804}
  (\bibinfo {year} {1988})}\BibitemShut {NoStop}%
\bibitem [{\citenamefont {Maeno}\ \emph {et~al.}()\citenamefont {Maeno},
  \citenamefont {Hashimoto}, \citenamefont {Yoshida}, \citenamefont
  {Nishizaki}, \citenamefont {Fujita}, \citenamefont {Bednorz},\ and\
  \citenamefont {Lichtenberg}}]{nature1994}%
  \BibitemOpen
  \bibfield  {author} {\bibinfo {author} {\bibfnamefont {Y.}~\bibnamefont
  {Maeno}}, \bibinfo {author} {\bibfnamefont {H.}~\bibnamefont {Hashimoto}},
  \bibinfo {author} {\bibfnamefont {K.}~\bibnamefont {Yoshida}}, \bibinfo
  {author} {\bibfnamefont {S.}~\bibnamefont {Nishizaki}}, \bibinfo {author}
  {\bibfnamefont {T.}~\bibnamefont {Fujita}}, \bibinfo {author} {\bibfnamefont
  {J.~G.}\ \bibnamefont {Bednorz}}, \ and\ \bibinfo {author} {\bibfnamefont
  {F.}~\bibnamefont {Lichtenberg}},\ }\href {\doibase 10.1038/372532a0} {\
  10.1038/372532a0}\BibitemShut {NoStop}%
\bibitem [{\citenamefont {Fu}\ and\ \citenamefont
  {Kane}(2008)}]{PhysRevLett.100.096407}%
  \BibitemOpen
  \bibfield  {author} {\bibinfo {author} {\bibfnamefont {L.}~\bibnamefont
  {Fu}}\ and\ \bibinfo {author} {\bibfnamefont {C.~L.}\ \bibnamefont {Kane}},\
  }\href {\doibase 10.1103/PhysRevLett.100.096407} {\bibfield  {journal}
  {\bibinfo  {journal} {Phys. Rev. Lett.}\ }\textbf {\bibinfo {volume} {100}},\
  \bibinfo {pages} {096407} (\bibinfo {year} {2008})}\BibitemShut {NoStop}%
\bibitem [{\citenamefont {Xu}\ \emph {et~al.}(2015)\citenamefont {Xu},
  \citenamefont {Wang}, \citenamefont {Liu}, \citenamefont {Ge}, \citenamefont
  {Yang}, \citenamefont {Liu}, \citenamefont {Xu}, \citenamefont {Guan},
  \citenamefont {Gao}, \citenamefont {Qian}, \citenamefont {Liu}, \citenamefont
  {Wang}, \citenamefont {Zhang}, \citenamefont {Xue},\ and\ \citenamefont
  {Jia}}]{PhysRevLett.114.017001}%
  \BibitemOpen
  \bibfield  {author} {\bibinfo {author} {\bibfnamefont {J.-P.}\ \bibnamefont
  {Xu}}, \bibinfo {author} {\bibfnamefont {M.-X.}\ \bibnamefont {Wang}},
  \bibinfo {author} {\bibfnamefont {Z.~L.}\ \bibnamefont {Liu}}, \bibinfo
  {author} {\bibfnamefont {J.-F.}\ \bibnamefont {Ge}}, \bibinfo {author}
  {\bibfnamefont {X.}~\bibnamefont {Yang}}, \bibinfo {author} {\bibfnamefont
  {C.}~\bibnamefont {Liu}}, \bibinfo {author} {\bibfnamefont {Z.~A.}\
  \bibnamefont {Xu}}, \bibinfo {author} {\bibfnamefont {D.}~\bibnamefont
  {Guan}}, \bibinfo {author} {\bibfnamefont {C.~L.}\ \bibnamefont {Gao}},
  \bibinfo {author} {\bibfnamefont {D.}~\bibnamefont {Qian}}, \bibinfo {author}
  {\bibfnamefont {Y.}~\bibnamefont {Liu}}, \bibinfo {author} {\bibfnamefont
  {Q.-H.}\ \bibnamefont {Wang}}, \bibinfo {author} {\bibfnamefont {F.-C.}\
  \bibnamefont {Zhang}}, \bibinfo {author} {\bibfnamefont {Q.-K.}\ \bibnamefont
  {Xue}}, \ and\ \bibinfo {author} {\bibfnamefont {J.-F.}\ \bibnamefont
  {Jia}},\ }\href {\doibase 10.1103/PhysRevLett.114.017001} {\bibfield
  {journal} {\bibinfo  {journal} {Phys. Rev. Lett.}\ }\textbf {\bibinfo
  {volume} {114}},\ \bibinfo {pages} {017001} (\bibinfo {year}
  {2015})}\BibitemShut {NoStop}%
\bibitem [{\citenamefont {Mourik}\ \emph {et~al.}(2012)\citenamefont {Mourik},
  \citenamefont {Zuo}, \citenamefont {Frolov}, \citenamefont {Plissard},
  \citenamefont {Bakkers},\ and\ \citenamefont {Kouwenhoven}}]{Mourik1003}%
  \BibitemOpen
  \bibfield  {author} {\bibinfo {author} {\bibfnamefont {V.}~\bibnamefont
  {Mourik}}, \bibinfo {author} {\bibfnamefont {K.}~\bibnamefont {Zuo}},
  \bibinfo {author} {\bibfnamefont {S.~M.}\ \bibnamefont {Frolov}}, \bibinfo
  {author} {\bibfnamefont {S.~R.}\ \bibnamefont {Plissard}}, \bibinfo {author}
  {\bibfnamefont {E.~P. A.~M.}\ \bibnamefont {Bakkers}}, \ and\ \bibinfo
  {author} {\bibfnamefont {L.~P.}\ \bibnamefont {Kouwenhoven}},\ }\href
  {\doibase 10.1126/science.1222360} {\bibfield  {journal} {\bibinfo  {journal}
  {Science}\ }\textbf {\bibinfo {volume} {336}},\ \bibinfo {pages} {1003}
  (\bibinfo {year} {2012})}\BibitemShut {NoStop}%
\bibitem [{\citenamefont {Deng}\ \emph {et~al.}(2012)\citenamefont {Deng},
  \citenamefont {Yu}, \citenamefont {Huang}, \citenamefont {Larsson},
  \citenamefont {Caroff},\ and\ \citenamefont {Xu}}]{doi:10.1021/nl303758w}%
  \BibitemOpen
  \bibfield  {author} {\bibinfo {author} {\bibfnamefont {M.~T.}\ \bibnamefont
  {Deng}}, \bibinfo {author} {\bibfnamefont {C.~L.}\ \bibnamefont {Yu}},
  \bibinfo {author} {\bibfnamefont {G.~Y.}\ \bibnamefont {Huang}}, \bibinfo
  {author} {\bibfnamefont {M.}~\bibnamefont {Larsson}}, \bibinfo {author}
  {\bibfnamefont {P.}~\bibnamefont {Caroff}}, \ and\ \bibinfo {author}
  {\bibfnamefont {H.~Q.}\ \bibnamefont {Xu}},\ }\href {\doibase
  10.1021/nl303758w} {\bibfield  {journal} {\bibinfo  {journal} {Nano Letters}\
  }\textbf {\bibinfo {volume} {12}},\ \bibinfo {pages} {6414} (\bibinfo {year}
  {2012})}\BibitemShut {NoStop}%
\bibitem [{\citenamefont {Albrecht}\ \emph {et~al.}(2016)\citenamefont
  {Albrecht}, \citenamefont {Higginbotham}, \citenamefont {Madsen},
  \citenamefont {Kuemmeth}, \citenamefont {Jespersen}, \citenamefont
  {Nyg{\aa}rd}, \citenamefont {Krogstrup},\ and\ \citenamefont
  {Marcus}}]{nat_charlie_markus}%
  \BibitemOpen
  \bibfield  {author} {\bibinfo {author} {\bibfnamefont {S.~M.}\ \bibnamefont
  {Albrecht}}, \bibinfo {author} {\bibfnamefont {A.~P.}\ \bibnamefont
  {Higginbotham}}, \bibinfo {author} {\bibfnamefont {M.}~\bibnamefont
  {Madsen}}, \bibinfo {author} {\bibfnamefont {F.}~\bibnamefont {Kuemmeth}},
  \bibinfo {author} {\bibfnamefont {T.~S.}\ \bibnamefont {Jespersen}}, \bibinfo
  {author} {\bibfnamefont {J.}~\bibnamefont {Nyg{\aa}rd}}, \bibinfo {author}
  {\bibfnamefont {P.}~\bibnamefont {Krogstrup}}, \ and\ \bibinfo {author}
  {\bibfnamefont {C.~M.}\ \bibnamefont {Marcus}},\ }\href
  {http://dx.doi.org/10.1038/nature17162} {\bibfield  {journal} {\bibinfo
  {journal} {Nature}\ }\textbf {\bibinfo {volume} {531}},\ \bibinfo {pages}
  {206} (\bibinfo {year} {2016})}\BibitemShut {NoStop}%
\bibitem [{\citenamefont {Nadj-Perge}\ \emph {et~al.}(2014)\citenamefont
  {Nadj-Perge}, \citenamefont {Drozdov}, \citenamefont {Li}, \citenamefont
  {Chen}, \citenamefont {Jeon}, \citenamefont {Seo}, \citenamefont {MacDonald},
  \citenamefont {Bernevig},\ and\ \citenamefont {Yazdani}}]{Yazdani}%
  \BibitemOpen
  \bibfield  {author} {\bibinfo {author} {\bibfnamefont {S.}~\bibnamefont
  {Nadj-Perge}}, \bibinfo {author} {\bibfnamefont {I.~K.}\ \bibnamefont
  {Drozdov}}, \bibinfo {author} {\bibfnamefont {J.}~\bibnamefont {Li}},
  \bibinfo {author} {\bibfnamefont {H.}~\bibnamefont {Chen}}, \bibinfo {author}
  {\bibfnamefont {S.}~\bibnamefont {Jeon}}, \bibinfo {author} {\bibfnamefont
  {J.}~\bibnamefont {Seo}}, \bibinfo {author} {\bibfnamefont {A.~H.}\
  \bibnamefont {MacDonald}}, \bibinfo {author} {\bibfnamefont {B.~A.}\
  \bibnamefont {Bernevig}}, \ and\ \bibinfo {author} {\bibfnamefont
  {A.}~\bibnamefont {Yazdani}},\ }\href {\doibase 10.1126/science.1259327}
  {\bibfield  {journal} {\bibinfo  {journal} {Science}\ }\textbf {\bibinfo
  {volume} {346}},\ \bibinfo {pages} {602} (\bibinfo {year}
  {2014})}\BibitemShut {NoStop}%
\bibitem [{\citenamefont {Wang}\ \emph {et~al.}(2018)\citenamefont {Wang},
  \citenamefont {Kong}, \citenamefont {Fan}, \citenamefont {Chen},
  \citenamefont {Zhu}, \citenamefont {Liu}, \citenamefont {Cao}, \citenamefont
  {Sun}, \citenamefont {Du}, \citenamefont {Schneeloch}, \citenamefont {Zhong},
  \citenamefont {Gu}, \citenamefont {Fu}, \citenamefont {Ding},\ and\
  \citenamefont {Gao}}]{Wang2017}%
  \BibitemOpen
  \bibfield  {author} {\bibinfo {author} {\bibfnamefont {D.}~\bibnamefont
  {Wang}}, \bibinfo {author} {\bibfnamefont {L.}~\bibnamefont {Kong}}, \bibinfo
  {author} {\bibfnamefont {P.}~\bibnamefont {Fan}}, \bibinfo {author}
  {\bibfnamefont {H.}~\bibnamefont {Chen}}, \bibinfo {author} {\bibfnamefont
  {S.}~\bibnamefont {Zhu}}, \bibinfo {author} {\bibfnamefont {W.}~\bibnamefont
  {Liu}}, \bibinfo {author} {\bibfnamefont {L.}~\bibnamefont {Cao}}, \bibinfo
  {author} {\bibfnamefont {Y.}~\bibnamefont {Sun}}, \bibinfo {author}
  {\bibfnamefont {S.}~\bibnamefont {Du}}, \bibinfo {author} {\bibfnamefont
  {J.}~\bibnamefont {Schneeloch}}, \bibinfo {author} {\bibfnamefont
  {R.}~\bibnamefont {Zhong}}, \bibinfo {author} {\bibfnamefont
  {G.}~\bibnamefont {Gu}}, \bibinfo {author} {\bibfnamefont {L.}~\bibnamefont
  {Fu}}, \bibinfo {author} {\bibfnamefont {H.}~\bibnamefont {Ding}}, \ and\
  \bibinfo {author} {\bibfnamefont {H.-J.}\ \bibnamefont {Gao}},\ }\href
  {\doibase 10.1126/science.aao1797} {\bibfield  {journal} {\bibinfo  {journal}
  {Science}\ } (\bibinfo {year} {2018}),\ 10.1126/science.aao1797}\BibitemShut
  {NoStop}%
\bibitem [{\citenamefont {He}\ \emph {et~al.}(2017)\citenamefont {He},
  \citenamefont {Pan}, \citenamefont {Stern}, \citenamefont {Burks},
  \citenamefont {Che}, \citenamefont {Yin}, \citenamefont {Wang}, \citenamefont
  {Lian}, \citenamefont {Zhou}, \citenamefont {Choi}, \citenamefont {Murata},
  \citenamefont {Kou}, \citenamefont {Chen}, \citenamefont {Nie}, \citenamefont
  {Shao}, \citenamefont {Fan}, \citenamefont {Zhang}, \citenamefont {Liu},
  \citenamefont {Xia},\ and\ \citenamefont {Wang}}]{He2017}%
  \BibitemOpen
  \bibfield  {author} {\bibinfo {author} {\bibfnamefont {Q.~L.}\ \bibnamefont
  {He}}, \bibinfo {author} {\bibfnamefont {L.}~\bibnamefont {Pan}}, \bibinfo
  {author} {\bibfnamefont {A.~L.}\ \bibnamefont {Stern}}, \bibinfo {author}
  {\bibfnamefont {E.~C.}\ \bibnamefont {Burks}}, \bibinfo {author}
  {\bibfnamefont {X.}~\bibnamefont {Che}}, \bibinfo {author} {\bibfnamefont
  {G.}~\bibnamefont {Yin}}, \bibinfo {author} {\bibfnamefont {J.}~\bibnamefont
  {Wang}}, \bibinfo {author} {\bibfnamefont {B.}~\bibnamefont {Lian}}, \bibinfo
  {author} {\bibfnamefont {Q.}~\bibnamefont {Zhou}}, \bibinfo {author}
  {\bibfnamefont {E.~S.}\ \bibnamefont {Choi}}, \bibinfo {author}
  {\bibfnamefont {K.}~\bibnamefont {Murata}}, \bibinfo {author} {\bibfnamefont
  {X.}~\bibnamefont {Kou}}, \bibinfo {author} {\bibfnamefont {Z.}~\bibnamefont
  {Chen}}, \bibinfo {author} {\bibfnamefont {T.}~\bibnamefont {Nie}}, \bibinfo
  {author} {\bibfnamefont {Q.}~\bibnamefont {Shao}}, \bibinfo {author}
  {\bibfnamefont {Y.}~\bibnamefont {Fan}}, \bibinfo {author} {\bibfnamefont
  {S.~C.}\ \bibnamefont {Zhang}}, \bibinfo {author} {\bibfnamefont
  {K.}~\bibnamefont {Liu}}, \bibinfo {author} {\bibfnamefont {J.}~\bibnamefont
  {Xia}}, \ and\ \bibinfo {author} {\bibfnamefont {K.~L.}\ \bibnamefont
  {Wang}},\ }\href {\doibase 10.1126/science.aag2792} {\bibfield  {journal}
  {\bibinfo  {journal} {Science}\ }\textbf {\bibinfo {volume} {357}},\ \bibinfo
  {pages} {294} (\bibinfo {year} {2017})}\BibitemShut {NoStop}%
\bibitem [{\citenamefont {Alicea}(2010)}]{PhysRevB.81.125318}%
  \BibitemOpen
  \bibfield  {author} {\bibinfo {author} {\bibfnamefont {J.}~\bibnamefont
  {Alicea}},\ }\href {\doibase 10.1103/PhysRevB.81.125318} {\bibfield
  {journal} {\bibinfo  {journal} {Phys. Rev. B}\ }\textbf {\bibinfo {volume}
  {81}},\ \bibinfo {pages} {125318} (\bibinfo {year} {2010})}\BibitemShut
  {NoStop}%
\bibitem [{\citenamefont {Lutchyn}\ \emph {et~al.}(2010)\citenamefont
  {Lutchyn}, \citenamefont {Sau},\ and\ \citenamefont
  {Das~Sarma}}]{PhysRevLett.105.077001}%
  \BibitemOpen
  \bibfield  {author} {\bibinfo {author} {\bibfnamefont {R.~M.}\ \bibnamefont
  {Lutchyn}}, \bibinfo {author} {\bibfnamefont {J.~D.}\ \bibnamefont {Sau}}, \
  and\ \bibinfo {author} {\bibfnamefont {S.}~\bibnamefont {Das~Sarma}},\ }\href
  {\doibase 10.1103/PhysRevLett.105.077001} {\bibfield  {journal} {\bibinfo
  {journal} {Phys. Rev. Lett.}\ }\textbf {\bibinfo {volume} {105}},\ \bibinfo
  {pages} {077001} (\bibinfo {year} {2010})}\BibitemShut {NoStop}%
\bibitem [{\citenamefont {Zareapour}\ \emph {et~al.}(2012)\citenamefont
  {Zareapour}, \citenamefont {Hayat}, \citenamefont {Zhao}, \citenamefont
  {Kreshchuk}, \citenamefont {Jain}, \citenamefont {Kwok}, \citenamefont {Lee},
  \citenamefont {Cheong}, \citenamefont {Xu}, \citenamefont {Yang},
  \citenamefont {Gu}, \citenamefont {Jia}, \citenamefont {Cava},\ and\
  \citenamefont {Burch}}]{Zareapour2012}%
  \BibitemOpen
  \bibfield  {author} {\bibinfo {author} {\bibfnamefont {P.}~\bibnamefont
  {Zareapour}}, \bibinfo {author} {\bibfnamefont {A.}~\bibnamefont {Hayat}},
  \bibinfo {author} {\bibfnamefont {S.~Y.~F.}\ \bibnamefont {Zhao}}, \bibinfo
  {author} {\bibfnamefont {M.}~\bibnamefont {Kreshchuk}}, \bibinfo {author}
  {\bibfnamefont {A.}~\bibnamefont {Jain}}, \bibinfo {author} {\bibfnamefont
  {D.~C.}\ \bibnamefont {Kwok}}, \bibinfo {author} {\bibfnamefont
  {N.}~\bibnamefont {Lee}}, \bibinfo {author} {\bibfnamefont {S.-W.}\
  \bibnamefont {Cheong}}, \bibinfo {author} {\bibfnamefont {Z.}~\bibnamefont
  {Xu}}, \bibinfo {author} {\bibfnamefont {A.}~\bibnamefont {Yang}}, \bibinfo
  {author} {\bibfnamefont {G.}~\bibnamefont {Gu}}, \bibinfo {author}
  {\bibfnamefont {S.}~\bibnamefont {Jia}}, \bibinfo {author} {\bibfnamefont
  {R.~J.}\ \bibnamefont {Cava}}, \ and\ \bibinfo {author} {\bibfnamefont
  {K.~S.}\ \bibnamefont {Burch}},\ }\href {\doibase 10.1038/ncomms2042}
  {\bibfield  {journal} {\bibinfo  {journal} {Nature Communications}\ }\textbf
  {\bibinfo {volume} {3}},\ \bibinfo {pages} {1056} (\bibinfo {year}
  {2012})}\BibitemShut {NoStop}%
\bibitem [{\citenamefont {Trani}\ \emph {et~al.}(2016)\citenamefont {Trani},
  \citenamefont {Campagnano}, \citenamefont {Tagliacozzo},\ and\ \citenamefont
  {Lucignano}}]{PhysRevB.94.134518}%
  \BibitemOpen
  \bibfield  {author} {\bibinfo {author} {\bibfnamefont {F.}~\bibnamefont
  {Trani}}, \bibinfo {author} {\bibfnamefont {G.}~\bibnamefont {Campagnano}},
  \bibinfo {author} {\bibfnamefont {A.}~\bibnamefont {Tagliacozzo}}, \ and\
  \bibinfo {author} {\bibfnamefont {P.}~\bibnamefont {Lucignano}},\ }\href
  {\doibase 10.1103/PhysRevB.94.134518} {\bibfield  {journal} {\bibinfo
  {journal} {Phys. Rev. B}\ }\textbf {\bibinfo {volume} {94}},\ \bibinfo
  {pages} {134518} (\bibinfo {year} {2016})}\BibitemShut {NoStop}%
\bibitem [{\citenamefont {Li}\ \emph {et~al.}(2015)\citenamefont {Li},
  \citenamefont {Chan},\ and\ \citenamefont {Yao}}]{PhysRevB.91.235143}%
  \BibitemOpen
  \bibfield  {author} {\bibinfo {author} {\bibfnamefont {Z.-X.}\ \bibnamefont
  {Li}}, \bibinfo {author} {\bibfnamefont {C.}~\bibnamefont {Chan}}, \ and\
  \bibinfo {author} {\bibfnamefont {H.}~\bibnamefont {Yao}},\ }\href {\doibase
  10.1103/PhysRevB.91.235143} {\bibfield  {journal} {\bibinfo  {journal} {Phys.
  Rev. B}\ }\textbf {\bibinfo {volume} {91}},\ \bibinfo {pages} {235143}
  (\bibinfo {year} {2015})}\BibitemShut {NoStop}%
\bibitem [{\citenamefont {Yan}\ \emph {et~al.}(2018)\citenamefont {Yan},
  \citenamefont {Song},\ and\ \citenamefont {Wang}}]{1803.08545}%
  \BibitemOpen
  \bibfield  {author} {\bibinfo {author} {\bibfnamefont {Z.}~\bibnamefont
  {Yan}}, \bibinfo {author} {\bibfnamefont {F.}~\bibnamefont {Song}}, \ and\
  \bibinfo {author} {\bibfnamefont {Z.}~\bibnamefont {Wang}},\ }\href {\doibase
  10.1103/PhysRevLett.121.096803} {\bibfield  {journal} {\bibinfo  {journal}
  {Phys. Rev. Lett.}\ }\textbf {\bibinfo {volume} {121}},\ \bibinfo {pages}
  {096803} (\bibinfo {year} {2018})}\BibitemShut {NoStop}%
\bibitem [{\citenamefont {Bednorz}\ and\ \citenamefont
  {M{\"u}ller}(1986)}]{Bednorz1986}%
  \BibitemOpen
  \bibfield  {author} {\bibinfo {author} {\bibfnamefont {J.~G.}\ \bibnamefont
  {Bednorz}}\ and\ \bibinfo {author} {\bibfnamefont {K.~A.}\ \bibnamefont
  {M{\"u}ller}},\ }\href {\doibase 10.1007/BF01303701} {\bibfield  {journal}
  {\bibinfo  {journal} {Zeitschrift f{\"u}r Physik B Condensed Matter}\
  }\textbf {\bibinfo {volume} {64}},\ \bibinfo {pages} {189} (\bibinfo {year}
  {1986})}\BibitemShut {NoStop}%
\bibitem [{\citenamefont {Anderson}(1987)}]{ANDERSON1196}%
  \BibitemOpen
  \bibfield  {author} {\bibinfo {author} {\bibfnamefont {P.~W.}\ \bibnamefont
  {Anderson}},\ }\href {\doibase 10.1126/science.235.4793.1196} {\bibfield
  {journal} {\bibinfo  {journal} {Science}\ }\textbf {\bibinfo {volume}
  {235}},\ \bibinfo {pages} {1196} (\bibinfo {year} {1987})}\BibitemShut
  {NoStop}%
\bibitem [{\citenamefont {Tsuei}\ and\ \citenamefont
  {Kirtley}(2000)}]{RevModPhys.72.969}%
  \BibitemOpen
  \bibfield  {author} {\bibinfo {author} {\bibfnamefont {C.~C.}\ \bibnamefont
  {Tsuei}}\ and\ \bibinfo {author} {\bibfnamefont {J.~R.}\ \bibnamefont
  {Kirtley}},\ }\href {\doibase 10.1103/RevModPhys.72.969} {\bibfield
  {journal} {\bibinfo  {journal} {Rev. Mod. Phys.}\ }\textbf {\bibinfo {volume}
  {72}},\ \bibinfo {pages} {969} (\bibinfo {year} {2000})}\BibitemShut
  {NoStop}%
\bibitem [{\citenamefont {Wenger}\ and\ \citenamefont
  {\"Ostlund}(1993)}]{WengerPhysRevB.47.5977}%
  \BibitemOpen
  \bibfield  {author} {\bibinfo {author} {\bibfnamefont {F.}~\bibnamefont
  {Wenger}}\ and\ \bibinfo {author} {\bibfnamefont {S.}~\bibnamefont
  {\"Ostlund}},\ }\href {\doibase 10.1103/PhysRevB.47.5977} {\bibfield
  {journal} {\bibinfo  {journal} {Phys. Rev. B}\ }\textbf {\bibinfo {volume}
  {47}},\ \bibinfo {pages} {5977} (\bibinfo {year} {1993})}\BibitemShut
  {NoStop}%
\bibitem [{\citenamefont {Senthil}\ \emph {et~al.}(1999)\citenamefont
  {Senthil}, \citenamefont {Marston},\ and\ \citenamefont
  {Fisher}}]{PhysRevB.60.4245}%
  \BibitemOpen
  \bibfield  {author} {\bibinfo {author} {\bibfnamefont {T.}~\bibnamefont
  {Senthil}}, \bibinfo {author} {\bibfnamefont {J.~B.}\ \bibnamefont
  {Marston}}, \ and\ \bibinfo {author} {\bibfnamefont {M.~P.~A.}\ \bibnamefont
  {Fisher}},\ }\href {\doibase 10.1103/PhysRevB.60.4245} {\bibfield  {journal}
  {\bibinfo  {journal} {Phys. Rev. B}\ }\textbf {\bibinfo {volume} {60}},\
  \bibinfo {pages} {4245} (\bibinfo {year} {1999})}\BibitemShut {NoStop}%
\bibitem [{\citenamefont {Morita}\ and\ \citenamefont
  {Hatsugai}(2000)}]{PhysRevB.62.99}%
  \BibitemOpen
  \bibfield  {author} {\bibinfo {author} {\bibfnamefont {Y.}~\bibnamefont
  {Morita}}\ and\ \bibinfo {author} {\bibfnamefont {Y.}~\bibnamefont
  {Hatsugai}},\ }\href {\doibase 10.1103/PhysRevB.62.99} {\bibfield  {journal}
  {\bibinfo  {journal} {Phys. Rev. B}\ }\textbf {\bibinfo {volume} {62}},\
  \bibinfo {pages} {99} (\bibinfo {year} {2000})}\BibitemShut {NoStop}%
\bibitem [{\citenamefont {Read}\ and\ \citenamefont
  {Green}(2000)}]{PhysRevB.61.10267}%
  \BibitemOpen
  \bibfield  {author} {\bibinfo {author} {\bibfnamefont {N.}~\bibnamefont
  {Read}}\ and\ \bibinfo {author} {\bibfnamefont {D.}~\bibnamefont {Green}},\
  }\href {\doibase 10.1103/PhysRevB.61.10267} {\bibfield  {journal} {\bibinfo
  {journal} {Phys. Rev. B}\ }\textbf {\bibinfo {volume} {61}},\ \bibinfo
  {pages} {10267} (\bibinfo {year} {2000})}\BibitemShut {NoStop}%
\bibitem [{\citenamefont {Sato}\ \emph {et~al.}(2010)\citenamefont {Sato},
  \citenamefont {Takahashi},\ and\ \citenamefont {Fujimoto}}]{Sato2010}%
  \BibitemOpen
  \bibfield  {author} {\bibinfo {author} {\bibfnamefont {M.}~\bibnamefont
  {Sato}}, \bibinfo {author} {\bibfnamefont {Y.}~\bibnamefont {Takahashi}}, \
  and\ \bibinfo {author} {\bibfnamefont {S.}~\bibnamefont {Fujimoto}},\ }\href
  {\doibase 10.1103/PhysRevB.82.134521} {\bibfield  {journal} {\bibinfo
  {journal} {Physical Review B}\ }\textbf {\bibinfo {volume} {82}},\ \bibinfo
  {pages} {134521} (\bibinfo {year} {2010})}\BibitemShut {NoStop}%
\bibitem [{\citenamefont {Chern}(2016)}]{doi:10.1063/1.4961462}%
  \BibitemOpen
  \bibfield  {author} {\bibinfo {author} {\bibfnamefont {T.}~\bibnamefont
  {Chern}},\ }\href {\doibase 10.1063/1.4961462} {\bibfield  {journal}
  {\bibinfo  {journal} {AIP Advances}\ }\textbf {\bibinfo {volume} {6}},\
  \bibinfo {pages} {085211} (\bibinfo {year} {2016})}\BibitemShut {NoStop}%
\bibitem [{\citenamefont {Awoga}\ \emph {et~al.}(2017)\citenamefont {Awoga},
  \citenamefont {Bouhon},\ and\ \citenamefont
  {Black-Schaffer}}]{PhysRevB.96.014521}%
  \BibitemOpen
  \bibfield  {author} {\bibinfo {author} {\bibfnamefont {O.~A.}\ \bibnamefont
  {Awoga}}, \bibinfo {author} {\bibfnamefont {A.}~\bibnamefont {Bouhon}}, \
  and\ \bibinfo {author} {\bibfnamefont {A.~M.}\ \bibnamefont
  {Black-Schaffer}},\ }\href {\doibase 10.1103/PhysRevB.96.014521} {\bibfield
  {journal} {\bibinfo  {journal} {Phys. Rev. B}\ }\textbf {\bibinfo {volume}
  {96}},\ \bibinfo {pages} {014521} (\bibinfo {year} {2017})}\BibitemShut
  {NoStop}%
\bibitem [{\citenamefont {Kiesel}\ \emph {et~al.}(2013)\citenamefont {Kiesel},
  \citenamefont {Platt}, \citenamefont {Hanke},\ and\ \citenamefont
  {Thomale}}]{PhysRevLett.111.097001}%
  \BibitemOpen
  \bibfield  {author} {\bibinfo {author} {\bibfnamefont {M.~L.}\ \bibnamefont
  {Kiesel}}, \bibinfo {author} {\bibfnamefont {C.}~\bibnamefont {Platt}},
  \bibinfo {author} {\bibfnamefont {W.}~\bibnamefont {Hanke}}, \ and\ \bibinfo
  {author} {\bibfnamefont {R.}~\bibnamefont {Thomale}},\ }\href {\doibase
  10.1103/PhysRevLett.111.097001} {\bibfield  {journal} {\bibinfo  {journal}
  {Phys. Rev. Lett.}\ }\textbf {\bibinfo {volume} {111}},\ \bibinfo {pages}
  {097001} (\bibinfo {year} {2013})}\BibitemShut {NoStop}%
\bibitem [{\citenamefont {Liu}\ \emph {et~al.}(2013)\citenamefont {Liu},
  \citenamefont {Liu}, \citenamefont {Wu}, \citenamefont {Yang},\ and\
  \citenamefont {Yao}}]{PhysRevLett.111.066804}%
  \BibitemOpen
  \bibfield  {author} {\bibinfo {author} {\bibfnamefont {F.}~\bibnamefont
  {Liu}}, \bibinfo {author} {\bibfnamefont {C.-C.}\ \bibnamefont {Liu}},
  \bibinfo {author} {\bibfnamefont {K.}~\bibnamefont {Wu}}, \bibinfo {author}
  {\bibfnamefont {F.}~\bibnamefont {Yang}}, \ and\ \bibinfo {author}
  {\bibfnamefont {Y.}~\bibnamefont {Yao}},\ }\href {\doibase
  10.1103/PhysRevLett.111.066804} {\bibfield  {journal} {\bibinfo  {journal}
  {Phys. Rev. Lett.}\ }\textbf {\bibinfo {volume} {111}},\ \bibinfo {pages}
  {066804} (\bibinfo {year} {2013})}\BibitemShut {NoStop}%
\bibitem [{\citenamefont {Lin}\ \emph {et~al.}(2016)\citenamefont {Lin},
  \citenamefont {Maiti},\ and\ \citenamefont {Chubukov}}]{PhysRevB.94.064519}%
  \BibitemOpen
  \bibfield  {author} {\bibinfo {author} {\bibfnamefont {S.-Z.}\ \bibnamefont
  {Lin}}, \bibinfo {author} {\bibfnamefont {S.}~\bibnamefont {Maiti}}, \ and\
  \bibinfo {author} {\bibfnamefont {A.}~\bibnamefont {Chubukov}},\ }\href
  {\doibase 10.1103/PhysRevB.94.064519} {\bibfield  {journal} {\bibinfo
  {journal} {Phys. Rev. B}\ }\textbf {\bibinfo {volume} {94}},\ \bibinfo
  {pages} {064519} (\bibinfo {year} {2016})}\BibitemShut {NoStop}%
\bibitem [{\citenamefont {Zhou}\ and\ \citenamefont
  {Wang}(2008)}]{PhysRevLett.100.217002}%
  \BibitemOpen
  \bibfield  {author} {\bibinfo {author} {\bibfnamefont {S.}~\bibnamefont
  {Zhou}}\ and\ \bibinfo {author} {\bibfnamefont {Z.}~\bibnamefont {Wang}},\
  }\href {\doibase 10.1103/PhysRevLett.100.217002} {\bibfield  {journal}
  {\bibinfo  {journal} {Phys. Rev. Lett.}\ }\textbf {\bibinfo {volume} {100}},\
  \bibinfo {pages} {217002} (\bibinfo {year} {2008})}\BibitemShut {NoStop}%
\bibitem [{\citenamefont {Gong}\ \emph {et~al.}(2017)\citenamefont {Gong},
  \citenamefont {Kargarian}, \citenamefont {Stern}, \citenamefont {Yue},
  \citenamefont {Zhou}, \citenamefont {Jin}, \citenamefont {Galitski},
  \citenamefont {Yakovenko},\ and\ \citenamefont {Xia}}]{Gong2017}%
  \BibitemOpen
  \bibfield  {author} {\bibinfo {author} {\bibfnamefont {X.}~\bibnamefont
  {Gong}}, \bibinfo {author} {\bibfnamefont {M.}~\bibnamefont {Kargarian}},
  \bibinfo {author} {\bibfnamefont {A.}~\bibnamefont {Stern}}, \bibinfo
  {author} {\bibfnamefont {D.}~\bibnamefont {Yue}}, \bibinfo {author}
  {\bibfnamefont {H.}~\bibnamefont {Zhou}}, \bibinfo {author} {\bibfnamefont
  {X.}~\bibnamefont {Jin}}, \bibinfo {author} {\bibfnamefont {V.~M.}\
  \bibnamefont {Galitski}}, \bibinfo {author} {\bibfnamefont {V.~M.}\
  \bibnamefont {Yakovenko}}, \ and\ \bibinfo {author} {\bibfnamefont
  {J.}~\bibnamefont {Xia}},\ }\href {\doibase 10.1126/sciadv.1602579} {\
  \textbf {\bibinfo {volume} {3}} (\bibinfo {year} {2017}),\
  10.1126/sciadv.1602579}\BibitemShut {NoStop}%
\bibitem [{\citenamefont {Nandkishore}\ \emph {et~al.}(2014)\citenamefont
  {Nandkishore}, \citenamefont {Levitov},\ and\ \citenamefont
  {Chubukov}}]{natphys8}%
  \BibitemOpen
  \bibfield  {author} {\bibinfo {author} {\bibfnamefont {R.}~\bibnamefont
  {Nandkishore}}, \bibinfo {author} {\bibfnamefont {L.~S.}\ \bibnamefont
  {Levitov}}, \ and\ \bibinfo {author} {\bibfnamefont {A.~V.}\ \bibnamefont
  {Chubukov}},\ }\href {\doibase 10.1038/nphys2208} {\bibfield  {journal}
  {\bibinfo  {journal} {Nature Physics}\ }\textbf {\bibinfo {volume} {8}},\
  \bibinfo {pages} {1804} (\bibinfo {year} {2014})}\BibitemShut {NoStop}%
\bibitem [{\citenamefont {Zhang}\ \emph {et~al.}(2017)\citenamefont {Zhang},
  \citenamefont {Covaci},\ and\ \citenamefont {Milo\ifmmode \check{s}\else
  \v{s}\fi{}evi\ifmmode~\acute{c}\else \'{c}\fi{}}}]{PhysRevB.96.224512}%
  \BibitemOpen
  \bibfield  {author} {\bibinfo {author} {\bibfnamefont {L.-F.}\ \bibnamefont
  {Zhang}}, \bibinfo {author} {\bibfnamefont {L.}~\bibnamefont {Covaci}}, \
  and\ \bibinfo {author} {\bibfnamefont {M.~V.}\ \bibnamefont {Milo\ifmmode
  \check{s}\else \v{s}\fi{}evi\ifmmode~\acute{c}\else \'{c}\fi{}}},\ }\href
  {\doibase 10.1103/PhysRevB.96.224512} {\bibfield  {journal} {\bibinfo
  {journal} {Phys. Rev. B}\ }\textbf {\bibinfo {volume} {96}},\ \bibinfo
  {pages} {224512} (\bibinfo {year} {2017})}\BibitemShut {NoStop}%
\bibitem [{\citenamefont {Becerra}\ and\ \citenamefont {Milo\ifmmode
  \check{s}\else \v{s}\fi{}evi\ifmmode~\acute{c}\else
  \'{c}\fi{}}(2016)}]{PhysRevB.94.184517}%
  \BibitemOpen
  \bibfield  {author} {\bibinfo {author} {\bibfnamefont {V.~F.}\ \bibnamefont
  {Becerra}}\ and\ \bibinfo {author} {\bibfnamefont {M.~V.}\ \bibnamefont
  {Milo\ifmmode \check{s}\else \v{s}\fi{}evi\ifmmode~\acute{c}\else
  \'{c}\fi{}}},\ }\href {\doibase 10.1103/PhysRevB.94.184517} {\bibfield
  {journal} {\bibinfo  {journal} {Phys. Rev. B}\ }\textbf {\bibinfo {volume}
  {94}},\ \bibinfo {pages} {184517} (\bibinfo {year} {2016})}\BibitemShut
  {NoStop}%
\bibitem [{\citenamefont {Sato}\ and\ \citenamefont
  {Fujimoto}(2010)}]{PhysRevLett.105.217001}%
  \BibitemOpen
  \bibfield  {author} {\bibinfo {author} {\bibfnamefont {M.}~\bibnamefont
  {Sato}}\ and\ \bibinfo {author} {\bibfnamefont {S.}~\bibnamefont
  {Fujimoto}},\ }\href {\doibase 10.1103/PhysRevLett.105.217001} {\bibfield
  {journal} {\bibinfo  {journal} {Phys. Rev. Lett.}\ }\textbf {\bibinfo
  {volume} {105}},\ \bibinfo {pages} {217001} (\bibinfo {year}
  {2010})}\BibitemShut {NoStop}%
\bibitem [{\citenamefont {Ortiz}\ \emph {et~al.}(2018)\citenamefont {Ortiz},
  \citenamefont {Varona}, \citenamefont {Viyuela},\ and\ \citenamefont
  {Martin-Delgado}}]{PhysRevB.97.064501}%
  \BibitemOpen
  \bibfield  {author} {\bibinfo {author} {\bibfnamefont {L.}~\bibnamefont
  {Ortiz}}, \bibinfo {author} {\bibfnamefont {S.}~\bibnamefont {Varona}},
  \bibinfo {author} {\bibfnamefont {O.}~\bibnamefont {Viyuela}}, \ and\
  \bibinfo {author} {\bibfnamefont {M.~A.}\ \bibnamefont {Martin-Delgado}},\
  }\href {\doibase 10.1103/PhysRevB.97.064501} {\bibfield  {journal} {\bibinfo
  {journal} {Phys. Rev. B}\ }\textbf {\bibinfo {volume} {97}},\ \bibinfo
  {pages} {064501} (\bibinfo {year} {2018})}\BibitemShut {NoStop}%
\bibitem [{\citenamefont {Fukui}\ \emph {et~al.}(2012)\citenamefont {Fukui},
  \citenamefont {Shiozaki}, \citenamefont {Fujiwara},\ and\ \citenamefont
  {Fujimoto}}]{Fukui2012}%
  \BibitemOpen
  \bibfield  {author} {\bibinfo {author} {\bibfnamefont {T.}~\bibnamefont
  {Fukui}}, \bibinfo {author} {\bibfnamefont {K.}~\bibnamefont {Shiozaki}},
  \bibinfo {author} {\bibfnamefont {T.}~\bibnamefont {Fujiwara}}, \ and\
  \bibinfo {author} {\bibfnamefont {S.}~\bibnamefont {Fujimoto}},\ }\href
  {\doibase 10.1143/JPSJ.81.114602} {\bibfield  {journal} {\bibinfo  {journal}
  {Journal of the Physical Society of Japan}\ }\textbf {\bibinfo {volume}
  {81}},\ \bibinfo {pages} {114602} (\bibinfo {year} {2012})}\BibitemShut
  {NoStop}%
\bibitem [{\citenamefont {Chiu}\ \emph {et~al.}(2016)\citenamefont {Chiu},
  \citenamefont {Teo}, \citenamefont {Schnyder},\ and\ \citenamefont
  {Ryu}}]{RevModPhysRyu}%
  \BibitemOpen
  \bibfield  {author} {\bibinfo {author} {\bibfnamefont {C.-K.}\ \bibnamefont
  {Chiu}}, \bibinfo {author} {\bibfnamefont {J.~C.~Y.}\ \bibnamefont {Teo}},
  \bibinfo {author} {\bibfnamefont {A.~P.}\ \bibnamefont {Schnyder}}, \ and\
  \bibinfo {author} {\bibfnamefont {S.}~\bibnamefont {Ryu}},\ }\href {\doibase
  10.1103/RevModPhys.88.035005} {\bibfield  {journal} {\bibinfo  {journal}
  {Rev. Mod. Phys.}\ }\textbf {\bibinfo {volume} {88}},\ \bibinfo {pages}
  {035005} (\bibinfo {year} {2016})}\BibitemShut {NoStop}%
\bibitem [{\citenamefont {Sedlmayr}\ \emph {et~al.}(2017)\citenamefont
  {Sedlmayr}, \citenamefont {Kaladzhyan}, \citenamefont {Dutreix},\ and\
  \citenamefont {Bena}}]{PhysRevB.96.184516}%
  \BibitemOpen
  \bibfield  {author} {\bibinfo {author} {\bibfnamefont {N.}~\bibnamefont
  {Sedlmayr}}, \bibinfo {author} {\bibfnamefont {V.}~\bibnamefont
  {Kaladzhyan}}, \bibinfo {author} {\bibfnamefont {C.}~\bibnamefont {Dutreix}},
  \ and\ \bibinfo {author} {\bibfnamefont {C.}~\bibnamefont {Bena}},\ }\href
  {\doibase 10.1103/PhysRevB.96.184516} {\bibfield  {journal} {\bibinfo
  {journal} {Phys. Rev. B}\ }\textbf {\bibinfo {volume} {96}},\ \bibinfo
  {pages} {184516} (\bibinfo {year} {2017})}\BibitemShut {NoStop}%
\bibitem [{\citenamefont {Sato}\ and\ \citenamefont {Ando}(2017)}]{Sato2016}%
  \BibitemOpen
  \bibfield  {author} {\bibinfo {author} {\bibfnamefont {M.}~\bibnamefont
  {Sato}}\ and\ \bibinfo {author} {\bibfnamefont {Y.}~\bibnamefont {Ando}},\
  }\href {http://stacks.iop.org/0034-4885/80/i=7/a=076501} {\bibfield
  {journal} {\bibinfo  {journal} {Reports on Progress in Physics}\ }\textbf
  {\bibinfo {volume} {80}},\ \bibinfo {pages} {076501} (\bibinfo {year}
  {2017})}\BibitemShut {NoStop}%
\bibitem [{\citenamefont {Teo}\ and\ \citenamefont {Kane}(2010)}]{TeoKane2010}%
  \BibitemOpen
  \bibfield  {author} {\bibinfo {author} {\bibfnamefont {J.~C.~Y.}\
  \bibnamefont {Teo}}\ and\ \bibinfo {author} {\bibfnamefont {C.~L.}\
  \bibnamefont {Kane}},\ }\href {\doibase 10.1103/PhysRevB.82.115120}
  {\bibfield  {journal} {\bibinfo  {journal} {Phys. Rev. B}\ }\textbf {\bibinfo
  {volume} {82}},\ \bibinfo {pages} {115120} (\bibinfo {year}
  {2010})}\BibitemShut {NoStop}%
\bibitem [{\citenamefont {Fukui}\ \emph {et~al.}(2005)\citenamefont {Fukui},
  \citenamefont {Hatsugai},\ and\ \citenamefont {Suzuki}}]{Fukui2005}%
  \BibitemOpen
  \bibfield  {author} {\bibinfo {author} {\bibfnamefont {T.}~\bibnamefont
  {Fukui}}, \bibinfo {author} {\bibfnamefont {Y.}~\bibnamefont {Hatsugai}}, \
  and\ \bibinfo {author} {\bibfnamefont {H.}~\bibnamefont {Suzuki}},\ }\href
  {\doibase 10.1143/JPSJ.74.1674} {\bibfield  {journal} {\bibinfo  {journal}
  {Journal of the Physical Society of Japan}\ }\textbf {\bibinfo {volume}
  {74}},\ \bibinfo {pages} {1674} (\bibinfo {year} {2005})}\BibitemShut
  {NoStop}%
\bibitem [{\citenamefont {Resta}(2000)}]{Resta2000}%
  \BibitemOpen
  \bibfield  {author} {\bibinfo {author} {\bibfnamefont {R.}~\bibnamefont
  {Resta}},\ }\href {http://stacks.iop.org/0953-8984/12/i=9/a=201} {\bibfield
  {journal} {\bibinfo  {journal} {Journal of Physics: Condensed Matter}\
  }\textbf {\bibinfo {volume} {12}},\ \bibinfo {pages} {R107} (\bibinfo {year}
  {2000})}\BibitemShut {NoStop}%
\bibitem [{\citenamefont {Mong}\ and\ \citenamefont
  {Shivamoggi}(2011)}]{Mong2011}%
  \BibitemOpen
  \bibfield  {author} {\bibinfo {author} {\bibfnamefont {R.~S.~K.}\
  \bibnamefont {Mong}}\ and\ \bibinfo {author} {\bibfnamefont {V.}~\bibnamefont
  {Shivamoggi}},\ }\href {\doibase 10.1103/PhysRevB.83.125109} {\bibfield
  {journal} {\bibinfo  {journal} {Phys. Rev. B}\ }\textbf {\bibinfo {volume}
  {83}},\ \bibinfo {pages} {125109} (\bibinfo {year} {2011})}\BibitemShut
  {NoStop}%
\bibitem [{\citenamefont {Ryu}\ and\ \citenamefont
  {Hatsugai}(2002)}]{Hatsugai2002}%
  \BibitemOpen
  \bibfield  {author} {\bibinfo {author} {\bibfnamefont {S.}~\bibnamefont
  {Ryu}}\ and\ \bibinfo {author} {\bibfnamefont {Y.}~\bibnamefont {Hatsugai}},\
  }\href {\doibase 10.1103/PhysRevLett.89.077002} {\bibfield  {journal}
  {\bibinfo  {journal} {Phys. Rev. Lett.}\ }\textbf {\bibinfo {volume} {89}},\
  \bibinfo {pages} {077002} (\bibinfo {year} {2002})}\BibitemShut {NoStop}%
\bibitem [{\citenamefont {Milošević}\ and\ \citenamefont
  {Perali}(2015)}]{0953-2048-28-6-060201}%
  \BibitemOpen
  \bibfield  {author} {\bibinfo {author} {\bibfnamefont {M.~V.}\ \bibnamefont
  {Milošević}}\ and\ \bibinfo {author} {\bibfnamefont {A.}~\bibnamefont
  {Perali}},\ }\href {http://stacks.iop.org/0953-2048/28/i=6/a=060201}
  {\bibfield  {journal} {\bibinfo  {journal} {Superconductor Science and
  Technology}\ }\textbf {\bibinfo {volume} {28}},\ \bibinfo {pages} {060201}
  (\bibinfo {year} {2015})}\BibitemShut {NoStop}%
\bibitem [{\citenamefont {Cao}\ \emph {et~al.}(2018)\citenamefont {Cao},
  \citenamefont {Fatemi}, \citenamefont {Fang}, \citenamefont {Watanabe},
  \citenamefont {Taniguchi}, \citenamefont {Kaxiras},\ and\ \citenamefont
  {Jarillo-Herrero}}]{jarillo}%
  \BibitemOpen
  \bibfield  {author} {\bibinfo {author} {\bibfnamefont {Y.}~\bibnamefont
  {Cao}}, \bibinfo {author} {\bibfnamefont {V.}~\bibnamefont {Fatemi}},
  \bibinfo {author} {\bibfnamefont {S.}~\bibnamefont {Fang}}, \bibinfo {author}
  {\bibfnamefont {K.}~\bibnamefont {Watanabe}}, \bibinfo {author}
  {\bibfnamefont {T.}~\bibnamefont {Taniguchi}}, \bibinfo {author}
  {\bibfnamefont {E.}~\bibnamefont {Kaxiras}}, \ and\ \bibinfo {author}
  {\bibfnamefont {P.}~\bibnamefont {Jarillo-Herrero}},\ }\href {\doibase
  10.1038/nature26160} {\bibfield  {journal} {\bibinfo  {journal} {Nature}\
  }\textbf {\bibinfo {volume} {556}},\ \bibinfo {pages} {43} (\bibinfo {year}
  {2018})}\BibitemShut {NoStop}%
\bibitem [{\citenamefont {Isobe}\ \emph {et~al.}(2018)\citenamefont {Isobe},
  \citenamefont {Yuan},\ and\ \citenamefont {Fu}}]{Fu_conclusions}%
  \BibitemOpen
  \bibfield  {author} {\bibinfo {author} {\bibfnamefont {H.}~\bibnamefont
  {Isobe}}, \bibinfo {author} {\bibfnamefont {N.~F.~Q.}\ \bibnamefont {Yuan}},
  \ and\ \bibinfo {author} {\bibfnamefont {L.}~\bibnamefont {Fu}},\ }\href
  {\doibase 10.1103/PhysRevX.8.041041} {\bibfield  {journal} {\bibinfo
  {journal} {Phys. Rev. X}\ }\textbf {\bibinfo {volume} {8}},\ \bibinfo {pages}
  {041041} (\bibinfo {year} {2018})}\BibitemShut {NoStop}%
\end{thebibliography}%
